\documentclass[
  reprint,
  superscriptaddress,
  amsmath,amssymb,
  aps,
  prb, % Use 'prb' for Physical Review B
  letterpaper,
  floatfix
]{revtex4-2}

\usepackage{natbib}
\usepackage{siunitx}
\usepackage{array}
\usepackage{xcolor}
\usepackage{subfigure} 
\usepackage{amsmath} 
\usepackage{amsfonts}
\usepackage{amssymb}
\usepackage{graphicx}
\usepackage{hyperref}

\providecommand{\vect}[1]{{\boldsymbol{#1}}}

\usepackage{physics}

\begin{document} 
\title{Tunable Phonon-Driven Magnon Spin Currents in Altermagnets}

\author{Sofie Helene Ursin}
\affiliation{Department of Engineering Sciences, University of Agder, 4879 Grimstad, Norway} 
\author{Mathias Kläui}
\affiliation{Institute of Physics, Johannes Gutenberg University Mainz, Staudingerweg 7, 55128 Mainz, Germany} 
\affiliation{Center for Quantum Spintronics, Department of Physics, Norwegian University of Science and Technology, 7491 Trondheim, Norway} 
\author{Kjetil M. D. Hals}
\affiliation{Department of Engineering Sciences, University of Agder, 4879 Grimstad, Norway} 
\date{\today}

%%%%%%%%%%%%%%%%%%%%%%%%%%%%%%%%%%%%%%%%%%
\newcommand{\Kjetil}[1]{\textcolor{red}{#1}} 
% So that Kjetil can make comments
%%%%%%%%%%%%%%%%%%%%%%%%%%%%%%%%%%%%%%%%%%%%%%%%%%%%%%%%%%%%%%%%%%%%%%%%%%%%%%% 
\begin{abstract}
Altermagnets have recently attracted considerable interest due to their unique symmetry-governed spintronic properties. Here, we investigate phonon-induced magnon spin currents in a two-dimensional altermagnet. Starting from a microscopic theory of the coupled magnon–phonon system, we derive the nonequilibrium magnon distribution generated by selective phonon excitations. We show that the resulting spin currents exhibit a pronounced $d$-wave symmetry with respect to the phonon momentum. Moreover, the spin current along the altermagnetic directions can be completely reversed by tuning the phonon frequency. These findings establish altermagnets as promising platforms for realizing highly tunable, phonon-driven coherent terahertz magnon spin currents.
\end{abstract}

\maketitle 

%%%%%%%%%%%%%%%%%%%%%%%%%%%%%%%%%%%%%%%%%%%%%%%%%%%%%%%%%%%%%%%%%%%%%%%%%%%%%%
% Introduction: 
%%%%%%%%%%%%%%%%%%%%%%%%%%%%%%%%%%%%%%%%%%%%%%%%%%%%%%%%%%%%%%%%%%%%%%%%%%%%%%
Altermagnetism has recently emerged as a novel class of antiferromagnetically ordered spin systems in which crystal symmetry, in combination with exchange coupling, lifts the spin degeneracy of the electronic band structure~\cite{Smejkal:PRX2022a,Smejkal:PRX2022b}. Recent angle-resolved photoemission spectroscopy experiments have revealed pronounced band splitting, with the spin polarization of the electronic bands exhibiting $g$-wave~\cite{Krempasky:nature2024,Lee:prl2024,Osumi:prb2024,Hajlaoui:am2024,Reimers:nc2024,liu:prl2024,Zeng:as2024,Yang:nc2025,Li:cp2025,Lu:nl2025} and $d$-wave~\cite{Jiang:np2025,Zhang:np2025} symmetries in momentum space --- a key hallmark of altermagnetism. This mechanism enables efficient spin-current generation without requiring net magnetization or strong spin–orbit coupling (SOC), while preserving the key advantages of antiferromagnets for spintronic applications~\cite{Jungwirth:arxiv2025}, including the absence of stray fields and ultrafast terahertz (THz) spin dynamics. In addition, altermagnets host a variety of distinctive symmetry-governed phenomena, such as anomalous transport effects~\cite{Feng:ne2022,Das:nc2022,Betancourt:prl2023,Han:sa2024,Chu:prl2025,Badura:nc2025,Zhou:n2025,Han:pra2025,Ruales:2026,Ruales:am2025,Leraand:prb2025} without net magnetization, unconventional optical responses~\cite{Amin:n2024,Pan:prl2026}, and spin-splitter torques~\cite{Bai:prl2022,Karube:prl2022,Bose:ne2022}.

The distinctive symmetry of altermagnets also manifests in their collective spin excitations~\cite{Smejkal:prl2023,Brekke:prb2023,Cui:prb2023,Gomonay:npjs2024}. The magnon spectrum exhibits a momentum-dependent chiral band splitting, with degeneracies remaining only 
on a symmetry-governed submanifold. Furthermore, the chiral splitting displays an alternating symmetry across the Brillouin zone. This behavior contrasts sharply with conventional antiferromagnets, where the magnon modes are typically doubly degenerate in the absence of external magnetic fields or other symmetry breaking~\cite{Auerbach:book}. 
Recent inelastic neutron scattering experiments have directly observed this chiral splitting in altermagnets~\cite{Liu:prl2024,Sun:prl2025}.
\begin{figure}[ht] 
\centering 
\includegraphics[width=1\linewidth]{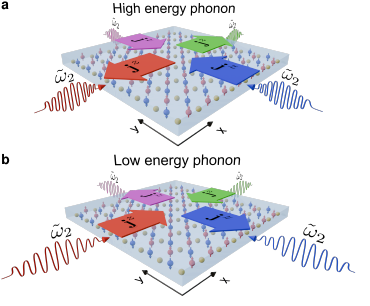}  
\caption{ Visualization of the complete reversal of the magnon spin current $\textbf{j}^z$ along the altermagnetic directions induced by tuning the phonon frequency $\tilde{\omega}_2(\vect{q})$. 
The colors associate selective phonon excitations with the resulting spin-current directions. 
(\textbf{a}) Spin currents generated by high-energy phonons. 
(\textbf{b}) Spin currents generated by low-energy phonons.}
\label{fig:3Dfigure}
\end{figure} 

The two nondegenerate chiral magnon branches in altermagnets carry opposite spin angular momentum, enabling magnon spin currents along altermagnetic directions without external magnetic fields. The resulting spin-current polarization is expected to show an alternating symmetry in the Brillouin zone, consistent with the chiral splitting. Experimental signatures of this alternating symmetry were recently observed by applying a temperature gradient along different crystallographic directions in the $d$-wave altermagnet LuFeO$_3$~\cite{Ruales:2026}.

An essential ingredient for understanding the generation and control of magnon spin currents is their coupling to the phonons of the underlying lattice. In altermagnets, this coupling is expected to play a particularly important role. 
First, phonons couple differently to the two spin sublattices as  the anisotropic exchange interaction give rise to different coupling strengths between phonons and the two magnon branches. 
Second, the band splitting along the altermagnetic directions leads to a momentum-dependent imbalance in the excitation of the two magnon branches. 
Despite its fundamental importance, the interplay between magnons and phonons in altermagnets has thus far received limited attention~\cite{Leraand:prb2026, Gunnink:prl2026, Bendin:arxiv}, and its consequences for magnon spin-current transport remain largely unexplored.

Here, we investigate the excitation of magnons by phonons in an altermagnetic Lieb lattice. By developing a microscopic theory of the coupled magnon–phonon system, we perturbatively compute—within the Keldysh formalism—the nonequilibrium magnon spin current induced by single coherent phonon modes. Remarkably, we find that the polarization of the generated spin current can be tuned via both the phonon frequency and momentum. For a fixed frequency, the spin-current polarization for different phonon propagation directions reflects the alternating symmetry of the chiral magnon band splitting. However, the overall sign of the spin current can be reversed by tuning the phonon frequency. Both the alternating symmetry and the frequency dependence of the spin current vanish in the absence of altermagnetic exchange coupling, identifying these effects as clear symmetry-governed signatures of altermagnetism. 

%%%%%%%%%%%%%%%%%%%%%%%%%%%%%%%%%%%%%%%%%%%%%%%%%%%%%%%%%%%%%%%%%%%%%%%%%%%%%%
% System: 
%%%%%%%%%%%%%%%%%%%%%%%%%%%%%%%%%%%%%%%%%%%%%%%%%%%%%%%%%%%%%%%%%%%%%%%%%%%%%% 
We consider a monolayer Lieb lattice with point group symmetry $D_{4h}$~\cite{Lieb_lattice_free}, grown on a substrate~\cite{Lieb_lattice_substrate}. The substrate breaks both inversion and the horizontal mirror symmetry, lowering the point group to $C_{4v}$. The resulting space group is the symmorphic group $P4mm$ and the corresponding magnetic point group contains the elements~\cite{Brekke:prb2023}:
$(E|0)$, $(C_{2z}|0)$, $(\sigma_{x}|0)$ $(\sigma_{y}|0)$, $(C_{4z}^+|\mathcal{T})$, $(C_{4z}^-|\mathcal{T})$, $(\sigma_{xy}|\mathcal{T})$, $(\sigma_{x\Bar{y}}|\mathcal{T})$.
Here, the operators on the left act on real space and the time-reversal operator $\mathcal{T}$ flips the spins. The symmetry operations $C_{2z}$ and $C_{4z}^{+}$ ($C_{4z}^{-}$) denote a twofold counterclockwise rotation and a fourfold counterclockwise (clockwise) rotation about the $z$ axis, respectively.  
$\sigma_{x}$, $\sigma_{y}$, $\sigma_{xy}$, $\sigma_{x\bar{y}}$ correspond to reflections with respect to the $xz$, $yz$, and the diagonal planes.
\begin{figure}[ht] 
\centering 
\includegraphics[width=1\linewidth]{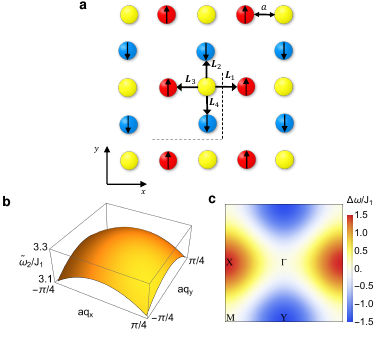}  
\caption{(\textbf{a}) A monolayer altermagnetic Lieb lattice. The unit cell contains three atoms: a spin-up, a spin-down, and a nonmagnetic atom. The four nearest-neighbor bonds connect each spin atom to the nonmagnetic atom, while no direct coupling occurs between the spin atoms.
(\textbf{b}) Phonon mode $\tilde{\omega}_2(\vect{q})=\omega_0^{(2)}\sqrt{1-a^2(q_x^2+q_y^2)/8}$ with $\omega_0^{(2)}= 1.3\omega_0$ where $\omega_0=4S\sqrt{K_z^2-4J_1K_z}$. (\textbf{c}) The magnon band-splitting $\Delta \omega(\vect{k})$ in Eq.~\eqref{eq:bandsplitting}. In (\textbf{b})-(\textbf{c}), we use $J_1= 25.7$~meV, $K_z/J_1=-0.1$, $\Bar{J}_2/J_1=-0.4$, $\Delta J_2/J_1=-0.2$, and $\hbar = S = 1$.}
\label{fig:aLiebLattice_bMagnonSplitting}
\end{figure} 

%%%%%%%%%%%%%%%%%%%%%%%%%%%%%%%%%%%%%%%%%%%%%%%%%%%%%%%%%%%%%%%%%%%%%%%%%%%%%%
% Phonon model: 
%%%%%%%%%%%%%%%%%%%%%%%%%%%%%%%%%%%%%%%%%%%%%%%%%%%%%%%%%%%%%%%%%%%%%%%%%%%%%% 
In the following, we investigate the nonequilibrium magnon spin current induced by selective phonon excitations. 
To this end, we first develop a theory for the phonons. 
Since our primary interest is the impact of individual coherent phonon modes, we treat a given mode as classically coherent in its coupling to the magnons. 
Motivated by laser-driven excitations, we focus on the optical phonon modes. 
To obtain analytic expressions for the phonon eigenvalues and eigenvectors, we restrict the analysis to nearest-neighbor interactions. 

The phonon spectrum is determined by the eigenvalue problem~\cite{Bruesch1982Phonons}
\begin{equation}
    \mathbf{D}(\vect{q}) \cdot \vect{e}_n(\vect{q}) = \tilde{\omega}^2_n(\vect{q})\vect{e}_n(\vect{q}),
\end{equation}
where
\begin{equation}
    \mathbf{D}(\vect{q}) = \begin{pmatrix}
        \mathbf{D}_{11} & \mathbf{D}_{12} & \mathbf{D}_{13} \\
        \mathbf{D}_{21} & \mathbf{D}_{22} & \mathbf{D}_{23} \\
        \mathbf{D}_{31} & \mathbf{D}_{32} & \mathbf{D}_{33} \\
    \end{pmatrix} \text{,\;\;  } \vect{e}_n(\vect{q}) = \begin{pmatrix}
        \vect{e}_n(1|\vect{q}) \\
        \vect{e}_n(2|\vect{q}) \\
        \vect{e}_n(3|\vect{q})
    \end{pmatrix},
\end{equation}
are the dynamical matrix and eigenvectors, respectively. Here, $\mathbf{D}_{kk'}$ is a $ 3\times3$ matrix with elements $D_{kk'}^{\alpha\beta}$ describing the interaction between atom $k$ and $k'$ in the unit cell. $\vect{e}_n(k|\vect{q})=\left[e_{n,x}(k|\vect{q}),e_{n,y}(k|\vect{q}),e_{n,z}(k|\vect{q})\right]^T$ is the part of the eigenvector corresponding to atom $k$ in the unit cell and $\tilde{\omega}_n(\vect{q})$ is the frequency of phonon mode $n$ with momentum $\vect{q}$.  We further label atom $1$ as spin-up, $2$ as spin-down and $3$ as the non-magnetic atom in the unit cell. Restricting to nearest-neighbor interactions, we obtain $\mathbf{D}_{12}=\mathbf{D}_{21}=\mathbf{0}$, since the spin atoms are not directly coupled. 
The remaining elements $\mathbf{D}_{kk'}$ are given by~\cite{Bruesch1982Phonons}
\begin{equation}
\label{eq:DynamicalMatrix}
D_{kk'}^{\alpha\beta}(\vect{q})=
\frac{1}{\sqrt{m_k m_{k'}}}
\sum_{l'}
\Phi_{\alpha\beta}
\left(
\begin{smallmatrix}
l & l' \\
k & k'
\end{smallmatrix}
\right)
e^{i \vect{q} \cdot 
\left[
\vect{r}
\left(
\begin{smallmatrix}
l' \\
k'
\end{smallmatrix}
\right)
-
\vect{r}
\left(
\begin{smallmatrix}
l \\
k
\end{smallmatrix}
\right)
\right]},
\end{equation}
where $m_k$ and $m_{k'}$ denote the atomic masses of sublattices $k$ and $k'$, respectively. 
The indices $l$ and $l'$ label the unit cell. 
The force constants $\Phi_{\alpha\beta}\left( \begin{smallmatrix} l & l' \\ k & k'\end{smallmatrix}\right)$ obey the symmetry constraints~\cite{Bruesch1982Phonons}
\begin{equation}
\vect{\Phi}(g\vect{L}) = 
\boldsymbol{S}\,\vect{\Phi}(\vect{L})\,\boldsymbol{S}^{T},
\label{eq:SymRels}
\end{equation}
where we have introduced the shorthand notation $\Phi_{\alpha\beta}(\vect{L}) \equiv \Phi_{\alpha\beta} \left( \begin{smallmatrix} l & l' \\ k & k' \end{smallmatrix} \right)$ with  
$\vect{L} \equiv \vect{r} \left( \begin{smallmatrix} l' \\ k' \end{smallmatrix} \right) - \vect{r} \left( \begin{smallmatrix} l \\ k \end{smallmatrix} \right)$,
for the elements of the $3\times3$ matrix $\vect{\Phi}(\vect{L})$. 
Here, $g \in P4mm$ denotes a space-group operation and $\boldsymbol{S} \in C_{4v}$ the corresponding $3\times3$ orthogonal matrix representation of the associated point-group operation~\cite{Comment1}.
In the present case, $\vect{\Phi}(\vect{L})$ is nonzero only for the nearest-neighbor bonds $\vect{L}_1$, $\vect{L}_2$, $\vect{L}_3$, and $\vect{L}_4$ shown in Fig.~\ref{fig:aLiebLattice_bMagnonSplitting}(a). 
To determine $\vect{\Phi}(\vect{L}_1)$, we identify the operations $g$ satisfying $g\vect{L}_1=\vect{L}_1$. 
The only symmetry elements that leave $\vect{L}_1$ invariant are $E$ and $\sigma_x$. 
Consequently, $\vect{\Phi}(\vect{L}_1) =  \boldsymbol{S}_{\sigma_x}\,\vect{\Phi}(\vect{L}_1)\,\boldsymbol{S}_{\sigma_x}^{T}$, from which we find the tensorial form 
\begin{equation}
\label{eq:Phi1}
\vect{\Phi}(\vect{L}_1)  =
- \begin{bmatrix}
        k_1 & 0 & k_4 \\
        0 & k_2 & 0 \\
        k_4' & 0 & k_3
    \end{bmatrix}   .  
 \end{equation}
The remaining matrices $\vect{\Phi}(\vect{L}_i)$ follow from successive fourfold rotations about the $z$-axis. 
Notably, the matrices are parametrized by five independent tensor coefficients: $k_1$, $k_2$, $k_3$, $k_4$, and $k_4'$.
The resulting force constants are then used to construct the dynamical matrix via Eq.~\eqref{eq:DynamicalMatrix}. For details, see the Supplemental Material (SM)~\cite{SuppMat}.

We obtain analytic expressions by diagonalizing $\mathbf{D}(\vect{q})$ perturbatively to second order in $\vect{q}$ near the $\Gamma$ point with $m_1=m_2=m_3/2$~\cite{SuppMat}.
This low-$\vect{q}$ expansion provides an accurate description of the phonon band over the range of magnon wave vectors $\vect{k}$ that predominantly contribute to the generated spin current~\cite{SuppMat}.
Below, we focus on the nondegenerate optical band [Fig.~\ref{fig:aLiebLattice_bMagnonSplitting}(b)]
\begin{equation}
\label{Eq:PhononBands}
\tilde{\omega}_2(\vect{q})= \omega_0^{(2)} \sqrt{ \left[1-a^2\left(q_x^2+q_y^2\right)/8 \right] },
\end{equation}
with the following corresponding sublattice eigenvectors:
$\vect{e}_2(1|\vect{q}) = [ ia_x^{(2)}q_x ,\, ia_y^{(2)} q_y ,\, \sin{\theta^{(2)}}\cos{\phi^{(2)}} ]$, 
$\vect{e}_2(2|\vect{q}) = [ ib_x^{(2)}q_x ,\, ib_y^{(2)}q_y ,\, \sin{\theta^{(2)}}\sin{\phi^{(2)}} ]$, 
$\vect{e}_2(3|\vect{q}) = [ ic_x^{(2)}q_x ,\, ic_y^{(2)}q_y ,\, \cos{\theta^{(2)}} ]$~\cite{comment2}.
Here, $\omega_0^{(2)}= \sqrt{4k_3/m_1}$, the parameters $a_i^{(2)}, b_i^{(2)}, c_i^{(2)}$ ($i=x,y,z$) depend on $k_1,k_2,k_3,k_4,k_4'$ and vanish in the limit $k_4,k_4' \to 0$~\cite{SuppMat}, and 
the angles are given by $\theta^{(2)}= -\pi/4$, and $\phi^{(2)}=\pi/4$.  
The vector $\vect{e}_2(\vect{q})$ is normalized to order $\vect{q}$.
The optical phonon~\eqref{Eq:PhononBands} carry an angular momentum of the form $\vect{J}\propto \vect{q}\times \hat{\boldsymbol{z}}$~\cite{SuppMat}. 
Since $\Gamma$ is a time-reversal-invariant momentum, the angular momentum vanishes as $\vect{q}\to 0$. 

%%%%%%%%%%%%%%%%%%%%%%%%%%%%%%%%%%%%%%%%%%%%%%%%%%%%%%%%%%%%%%%%%%%%%%%%%%%%%%
% Magnon model: 
%%%%%%%%%%%%%%%%%%%%%%%%%%%%%%%%%%%%%%%%%%%%%%%%%%%%%%%%%%%%%%%%%%%%%%%%%%%%%% 
The spin system is described by the Hamiltonian ($A$ and $B$ label the up and down spin lattice)~\cite{Smejkal:prl2023,Brekke:prb2023,Cui:prb2023}
\begin{equation}
H = H_1 + H_2 + H_2' + H_a,
\label{Eq:SpinH}
\end{equation}
where $H_1 = J_1 \sum_{\langle i,j \rangle} \vect{S}_i^A \!\cdot\! \vect{S}_j^B$ 
denotes an antiferromagnetic exchange interaction with $J_1>0$, $H_2 = J_2  (\sum_{\langle i_x,j_x \rangle} \vect{S}_i^A \!\cdot\! \vect{S}_j^A  + \sum_{\langle i_y,j_y \rangle} \vect{S}_i^B \!\cdot\! \vect{S}_j^B  )$ and  $H_2' = J_2'  (\sum_{\langle i_y,j_y \rangle} \vect{S}_i^A \!\cdot\! \vect{S}_j^A  + \sum_{\langle i_x,j_x \rangle} \vect{S}_i^B \!\cdot\! \vect{S}_j^B )$ represent ferromagnetic exchange interactions with $J_2,J_2'<0$, and 
$H_a = K_z \sum_i \!\left[(S_{i,z}^A)^2 + (S_{i,z}^B)^2\right] $ is an easy-axis anisotropy term with $K_z<0$.
$\langle i,j \rangle$, $\langle i_x,j_x \rangle$, and $\langle i_y,j_y \rangle$ denote nearest-neighbor bonds along the diagonal, $x$, and $y$ directions, respectively.
The collective spin excitations of Eq.~\eqref{Eq:SpinH} are obtained via a Holstein–Primakoff (HP) transformation~\cite{Auerbach:book}, resulting in the bosonic Bogoliubov–de Gennes (BdG) Hamiltonian
\begin{equation}
H_m = \sum_{\vect{k}} 
\begin{pmatrix} 
a_{\vect{k}}^\dag & b_{-\vect{k}} 
\end{pmatrix}
\begin{pmatrix}
A_{\vect{k}} & B_{\vect{k}} \\
B_{\vect{k}} & C_{\vect{k}}
\end{pmatrix}
\begin{pmatrix}
a_{\vect{k}} \\
b_{-\vect{k}}^\dag
\end{pmatrix}.
\end{equation}
Here, $A_{\vect{k}}$, $B_{\vect{k}}$, and $C_{\vect{k}}$ depend on $J_1$, $J_2$, $J_2'$, and $K_z$ (see SM~\cite{SuppMat}). 
To parameterize the altermagnetic splitting, we introduce $J_2=\bar{J}_2+\Delta J_2$ and $J_2'=\bar{J}_2-\Delta J_2$, such that $J_2-J_2'=2\Delta J_2$. 
$H_m$ is diagonalized by the BdG transformation
$a_{\vect{k}} = u_{\vect{k}}\alpha_{\vect{k}} + v_{\vect{k}}\beta_{-\vect{k}}^\dag$, 
$b_{-\vect{k}}^\dag = v_{\vect{k}}^*\alpha_{\vect{k}} + u_{\vect{k}}^*\beta_{-\vect{k}}^\dag$,
which yields
$H_m = \sum_{\vect{k}}  [ \omega_\alpha(\vect{k}) \alpha_{\vect{k}}^\dag \alpha_{\vect{k}}  +  \omega_\beta(\vect{k}) \beta_{\vect{k}}^\dag \beta_{\vect{k}} ] $
with eigenenergies $\omega_\alpha(\vect{k}) = (A_{\vect{k}} - C_{\vect{k}} + \sqrt{(A_{\vect{k}} + C_{\vect{k}})^2 - 4B_{\vect{k}}^2} )/2$, and
$\omega_\beta(\vect{k}) = (C_{\vect{k}} - A_{\vect{k}} + \sqrt{(A_{\vect{k}} + C_{\vect{k}})^2 - 4B_{\vect{k}}^2} )/2$.
The resulting band splitting $\Delta\omega(\vect{k}) = \omega_\alpha(\vect{k}) - \omega_\beta(\vect{k})$ is shown in Fig.~\ref{fig:aLiebLattice_bMagnonSplitting}(c) and assumes the form
\begin{equation}
\label{eq:bandsplitting}
\Delta\omega(\vect{k}) 
= 4S\Delta J_2 
\left[\cos(2k_x a) - \cos(2k_y a)\right].
\end{equation}
The characteristic $d$-wave symmetry is evident, yielding a finite spin splitting along the altermagnetic directions $\Gamma$–X and $\Gamma$–Y, and vanishing splitting along the symmetric direction $\Gamma$–M.

%%%%%%%%%%%%%%%%%%%%%%%%%%%%%%%%%%%%%%%%%%%%%%%%%%%%%%%%%%%%%%%%%%%%%%%%%%%%%%
% Magnon-Phonon coupling and spin current: 
%%%%%%%%%%%%%%%%%%%%%%%%%%%%%%%%%%%%%%%%%%%%%%%%%%%%%%%%%%%%%%%%%%%%%%%%%%%%%% 
The magnon–phonon coupling originates from the modulation of the exchange interactions by lattice vibrations. 
To leading order in the atomic displacements, the exchange interaction between sites $i$ and $j$ becomes $J_{ij} = J_{ij}^0 + \alpha_{ij}\,\hat{\vect{n}}_{ij} \cdot \Delta\vect{u}_{ij}$,
where $J_{ij}^0$ denotes the equilibrium exchange constant, $\alpha_{ij} = \partial J_{ij}/\partial a$ ($a$ is the lattice constant)~\cite{comment4} is the spin–phonon coupling parameter,  $\hat{\vect{n}}_{ij}$ is the unit vector pointing from site $i$ to $j$, and 
$\Delta\vect{u}_{ij} = \vect{u}_j - \vect{u}_i$ with $\vect{u}_i$ the displacement field of the atom at site $i$. 
Substituting this expansion into Eq.~\eqref{Eq:SpinH} yields the interaction Hamiltonian 
$H_I = H_{I1} + H_{I2} + H_{I3}$. 
The nearest-neighbor contribution reads $H_{I1} = \alpha_1 \sum_{\langle i,j \rangle} \hat{\vect{n}}_{ij}\!\cdot\!\Delta\vect{u}_{ij}\, \vect{S}_i^A \!\cdot\! \vect{S}_j^B$, with $\alpha_1 = \partial J_1/\partial a$. 
The next-nearest-neighbor terms are $H_{I2} = \sum_{\langle i_x,j_x \rangle} \hat{\vect{n}}_{ij}\!\cdot\!\Delta\vect{u}_{ij} [ \alpha_2 \vect{S}_i^A \!\cdot\! \vect{S}_j^A + \alpha_2' \vect{S}_i^B \!\cdot\! \vect{S}_j^B ]$ and 
$H_{I3} = \sum_{\langle i_y,j_y \rangle} \hat{\vect{n}}_{ij}\!\cdot\!\Delta\vect{u}_{ij} [ \alpha_2 \vect{S}_i^B \!\cdot\! \vect{S}_j^B + \alpha_2' \vect{S}_i^A \!\cdot\! \vect{S}_j^A ]$
where $\alpha_2 \equiv \bar{\alpha}_2 + \Delta\alpha_2$ and  $\alpha_2' \equiv \bar{\alpha}_2 - \Delta\alpha_2$ are given by the derivatives of $J_2$ and $J_2'$ with respect to $a$. 
The displacement field at sublattice site $k$ in unit cell $l$ can be expanded in the phonon eigenmodes as~\cite{Bruesch1982Phonons}
$\vect{u}  \left( \begin{smallmatrix} l \\ k \end{smallmatrix} \right) = \sqrt{1/N m_k} \sum_{n,\vect{q}}  \vect{e}_n(k|\vect{q}) Q_n(\vect{q},t) \exp (i\vect{q}\cdot \vect{r} \left( \begin{smallmatrix} l \\ k \end{smallmatrix} \right) )$
where $Q_n(\vect{q},t)$ denotes the normal coordinate and $N$ the number of unit cells. 
Upon inserting this expansion together with the HP representation of the spin operators into $H_I$ and performing a Fourier transformation to momentum space, we obtain the Hamiltonian~\cite{comment3}
\begin{equation}
\label{eq:H_I}
\begin{split}
    &H_I = \sum_{n,\vect{k},\vect{q}} \big[\Omega_{\Bar{\alpha}\alpha}(n,\vect{q},\vect{k})\alpha^\dag_{\vect{k}+\vect{q}}\alpha_{\vect{k}} +\Omega_{\Bar{\beta}\beta}(n,\vect{q},\vect{k})\beta^\dag_{-\vect{k}}\beta_{-\vect{k}-\vect{q}} \\
    &+ \Omega_{\Bar{\alpha}\Bar{\beta}}(n,\vect{q},\vect{k})\alpha^\dag_{\vect{k}+\vect{q}}\beta^\dag_{-\vect{k}} +\Omega_{\alpha\beta}(n,\vect{q},\vect{k})\alpha_{\vect{k}}\beta_{-\vect{k}-\vect{q}} \big] Q_n(\vect{q},t) ,
\end{split}
\end{equation}
where the coupling vertices $\Omega$ are given in the SM~\cite{SuppMat}.

%spin current:
Next, we examine the generation of altermagnonic dc spin currents driven by selective phonon excitations in the stationary regime. 
The nonequilibrium correction to the magnon distribution function $n_{\gamma}(\vect{k}) = \langle \gamma_{\vect{k}}^{\dagger}(t) \gamma_{\vect{k}}(t) \rangle$ arising from the coupling to an excited phonon mode $(n,\vect{q})$ can be expressed in terms of the magnons' Keldysh Green’s function
$G^{K}_{\gamma\bar{\gamma}}(\vect{k};\omega) $ as~\cite{Rammer_2007}
\begin{equation}
\label{eq:DistributionFunctionFormula}
\Delta n_{\gamma}(\vect{k})
=
\frac{i}{4\pi}
\int_{-\infty}^{\infty}
G^{K}_{\gamma\bar{\gamma}}(\vect{k};\omega)\, d\omega,
\end{equation}
where $\gamma \in \{\alpha,\beta\}$. $G^{K}$ is obtained from the contour-ordered Green’s function
$G_{\gamma\Bar{\gamma}}^{(\tau \tau' )}(\vect{k},t;\vect{k}',t') = -i\big\langle\mathcal{T}_c e^{-\frac{i}{{\hbar}}{\int_c}H_I(t_1)dt_1}\gamma_\vect{k}(t,\tau)\gamma^\dag_\vect{k}(t',\tau')\big\rangle_0$,
where the subscript $0$ denotes averaging over the equilibrium state in the infinite past, in which the system is noninteracting. 
The operator $\mathcal{T}_c$ enforces time ordering along the Keldysh contour $c$, whose upper/lower branch is labeled by $\tau$  and $\tau'$. 
All operators are taken in the interaction picture. 
The interaction Hamiltonian $H_I$ includes the contribution from a single phonon mode $(n,\vect{q})$, corresponding to a normal coordinate of the form~\cite{Bruesch1982Phonons}
$Q_{n'} (\vect{q}' , t)= (A_n(\vect{q}) /2) \delta_{nn'}\delta_{\vect{q},\vect{q}'} {\rm e}^{-i \tilde{\omega}_n ( {\vect{q}}) t} + (A_n^{\ast}(\vect{q})/ 2) \delta_{nn'}\delta_{-\vect{q},\vect{q}'} {\rm e}^{ i \tilde{\omega}_n ( {\vect{q}}) t} $, 
where $A_n(\vect{q})$ parametrizes the oscillation amplitude.
We evaluate  $G_{\gamma\bar{\gamma}}^{(\tau \tau' )}$  to second order in the magnon–phonon interaction. 
Expanding $G_{\gamma\Bar{\gamma}}^{(\tau \tau' )}$ to second order in $H_I$, applying Wick’s theorem, and transforming to frequency space and the triagonal representation in Keldysh space, we find the equation
\begin{equation}
\label{eq:Dyson}
     \underline{G}_{\gamma\Bar{\gamma}}(\vect{k};\omega)  = \underline{G}_{\gamma\Bar{\gamma}}(\vect{k};\omega)_0 +\underline{G}_{\gamma\Bar{\gamma}}(\vect{k};\omega)_0\underline{\Sigma}_{\gamma\Bar{\gamma}}(\vect{k};\omega)\underline{G}_{\gamma\Bar{\gamma}}(\vect{k};\omega)_0.
\end{equation}
with $\underline{G}=[G^{R},G^{K};0,G^{A}]$ and  $\underline{\Sigma}=[\Sigma^{R},\Sigma^{K};0,\Sigma^{A}]$. 
Here, $G^{R(A)}$ denotes the retarded (advanced) Green’s function, and the self-energy $\underline{\Sigma}$ is evaluated to second order in $H_I$ (see SM~\cite{SuppMat}). 
Extracting the Keldysh component of Eq.~\eqref{eq:Dyson} and substituting it into Eq.~\eqref{eq:DistributionFunctionFormula}, we obtain the following nonequilibrium correction to the magnon distribution function~\cite{SuppMat}
\begin{equation}
\label{eq:DistributionFunctions}
\begin{split}
    \Delta n_\gamma(\vect{k},\vect{q}) = &\sum_{\eta=\pm 1} \Bigg[\frac{K_{1\gamma} (n, \eta\vect{q}, \vect{k}) (2n_\gamma^0(\vect{k}+\eta\vect{q})+1)}{(\omega_\gamma(\vect{k}+\eta\vect{q})-\omega_\gamma(\vect{k})-\eta\tilde{ \omega}_n(\vect{q}))^2+\varepsilon^2} \\
    &+\frac{K_{2\gamma} (n, \eta\vect{q}, \vect{k}) (2n_{\underline{\gamma}}^0(-\vect{k}-\eta\vect{q})+1)}{(\omega_{\underline{\gamma}}(-\vect{k}-\eta\vect{q})+\omega_\gamma(\vect{k})+\eta \tilde{\omega}_n(\vect{q}))^2+\varepsilon^2}\Bigg], 
\end{split}
\end{equation}
where $\underline{\alpha}=\beta, \underline{\beta}=\alpha$, and $n_\gamma^0(\vect{k})$ is the Bose-Einstein distribution. The coefficients $K_{1\gamma}$ and $K_{2\gamma}$ depend on the coupling vertices $\Omega$ in Eq.~\eqref{eq:H_I} and can be found in the SM~\cite{SuppMat}. 
Additionally, we have included a phenomenological damping rate $\varepsilon$, approximated as $\varepsilon = \alpha_G \omega_\gamma(0)$ where $\alpha_G$ is the Gilbert damping parameter. 

\begin{figure}[htb]
    \includegraphics[width=1\linewidth]{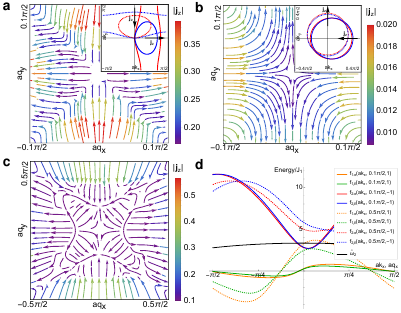}
    \caption{ Dimensionless spin current $\vect{j}^z(\vect{q})=(4\pi^2 a/J_1)\vect{J}^z(\vect{q})$ as a function of the phonon momentum $\vect{q}$ for three phonon spectra $\tilde{\omega}_2(\vect{q})=\omega_0^{(2)}\sqrt{1-a^2(q_x^2+q_y^2)/8}$: 
(\textbf{a}) $\omega_0^{(2)}=0.1\omega_0$, 
(\textbf{b}) $\omega_0^{(2)}=2.0\omega_0$, and 
(\textbf{c}) $\omega_0^{(2)}=1.3\omega_0$, where $\omega_0=4S\sqrt{K_z^2-4J_1K_z}$. 
The insets in (\textbf{a}) and (\textbf{b}) show the resonant $\vect{k}$ states governing the induced spin current. 
(\textbf{d}) Functions $f_{1\gamma}$ and $f_{2\gamma}$ versus $ak_x$ for $a\vect{q}=0.1\pi\hat{\vect{x}}/2$ (solid lines) and $a\vect{q}=0.5\pi\hat{\vect{x}}/2$ (dotted lines), together with $\tilde{\omega}_2(\vect{q})$ for $\omega_0^{(2)}=1.3\omega_0$. 
In all calculations, we use $J_1=25.7$~meV, $K_z/J_1=-0.1$, $\bar{J}_2/J_1=-0.4$, $\Delta J_2/J_1=-0.2$, $\alpha_1=-53$~meV/\AA, $\bar{\alpha}_2/\alpha_1=-0.4$, $\Delta\alpha_2/\alpha_1=-0.2$, $\alpha_G=10^{-1}$, $T=298$~K, $\hbar=S=1$, and $a=4$~\AA.
Furthermore, we chose $\sqrt{A_n^2/(N m_1 a^2)}=10^{-2}$, corresponding to a phonon amplitude of $4$~pm.}
    \label{fig:j=2SpinCurrent}
\end{figure}

Eq.~\eqref{eq:DistributionFunctions} allows us to identify the dominant processes contributing to the nonequilibrium distribution function. 
The first term describes single-magnon processes in which a magnon $\gamma$ either absorbs or emits a phonon with momentum $\vect{q}$ and energy $\tilde{\omega}_n(\vect{q})$. 
Energy conservation requires 
$\omega_\gamma(\vect{k}+\vect{q})=\omega_\gamma(\vect{k})+\tilde{\omega}_n(\vect{q})$ 
for absorption and 
$\omega_\gamma(\vect{k}-\vect{q})=\omega_\gamma(\vect{k})-\tilde{\omega}_n(\vect{q})$ 
for emission. 
The second term describes two-magnon processes in which magnons $\gamma$ and $\underline{\gamma}$ are either created (annihilated) accompanied by phonon annihilation (creation).
For small $\vect{q}$, one has $\omega_\gamma(\vect{k}+\eta\vect{q})\approx\omega_\gamma(\vect{k})$, implying that the single-magnon processes are resonant at small phonon energies. 
In contrast, the two-magnon processes require a comparatively large phonon energy due to the conservation of the energy of both magnons.
Consequently, the second term in Eq.~\eqref{eq:DistributionFunctions} can dominate at higher phonon energies. 
Importantly, as shown below, \emph{these two mechanisms generate phonon-driven magnon spin currents of opposite sign, enabling a reversal of the spin current by tuning the frequency of the excited phonon}.

The spin current polarized along $z$ and flowing in the direction $i \in \{x,y\}$ is calculated from the expression
\begin{equation}
\label{Eq:Jz}
    J^z_i(\vect{q}) = \frac{\hbar}{A}\sum_\vect{k}\left[v_i^\alpha(\vect{k}) \Delta n_\alpha(\vect{k},\vect{q})- v_i^\beta(\vect{k}) \Delta n_\beta(\vect{k},\vect{q})\right].
\end{equation}
Here, $A$ denotes the area of the sample and $v_i^\gamma(\vect{k}) = \partial \omega_\gamma(\vect{k})/\partial k_i$ 
is the group velocity. The summation runs over all $\vect{k}$ values within the first Brillouin zone (FBZ).
In the absence of the Dzyaloshinskii--Moriya interaction, $v_i^\gamma(\vect{k})$ contains no anomalous contribution arising from the magnon Berry curvature~\cite{Syljuaasen_prb2025}.

%%%%%%%%%%%%%%%%%%%%%%%%%%%%%%%%%%%%%%%%%%%%%%%%%%%%%%%%%%%%%%%%%%%%%%%%%%%%%%
% Results and discussion: 
%%%%%%%%%%%%%%%%%%%%%%%%%%%%%%%%%%%%%%%%%%%%%%%%%%%%%%%%%%%%%%%%%%%%%%%%%%%%%% 
Figs.~\ref{fig:j=2SpinCurrent}(a)-(b) show the spin current generated by the $n=2$ phonon mode in Eq.~\eqref{Eq:PhononBands} as a function of $\vect{q}$ for $\omega_0^{(2)}=0.1\omega_0$ and $\omega_0^{(2)}=2.0\omega_0$, respectively.
Here, $\omega_0=4S\sqrt{K_z^2-4J_1K_z}$. 
In both cases, the resulting spin current is strongly anisotropic and exhibits a pronounced $d$-wave symmetry, consistent with the anisotropic thermally driven magnon spin current reported in Ref.~\cite{Ruales:2026}. 
A key observation is the sign reversal of the spin current along the altermagnetic directions $[\pm1,0,0]$ and $[0,\pm1,0]$ when $\omega_0^{(2)}$ is changed from $0.1 \omega_0$ to $2.0 \omega_0$. 
This behavior can be understood by analyzing the dominant $\vect{k}$ states contributing to the spin current~\eqref{Eq:Jz}. 
These correspond to $\vect{k}\in\mathrm{FBZ}$ satisfying one of the resonance conditions
$f_{1\gamma}(\vect{k},\vect{q},\eta)=\tilde{\omega}_2(\vect{q})$ or
$f_{2\gamma}(\vect{k},\vect{q},\eta)=\tilde{\omega}_2(\vect{q})$, where
$f_{1\gamma}\equiv\omega_\gamma(\vect{k}+\eta\vect{q})-\omega_\gamma(\vect{k})$ and
$f_{2\gamma}\equiv\omega_{\underline{\gamma}}(-\vect{k}-\eta\vect{q})+\omega_\gamma(\vect{k})$
define the resonance conditions for single-magnon and two-magnon processes, respectively.
For $\omega_0^{(2)} = 0.1\omega_0$, single-magnon processes dominate; i.e., there exist $\vect{k}\in\mathrm{FBZ}$ satisfying $f_{1\gamma}(\vect{k},\vect{q},\eta)=\tilde{\omega}_2(\vect{q})$.
The corresponding resonant $\vect{k}$ states are shown in the inset of Fig.~\ref{fig:j=2SpinCurrent}(a) for $\vect{q}=(0.1\,\pi/2a)\,\hat{\vect{x}}$ (solid lines) and $\vect{q}=(0.1\,\pi/2a)\,\hat{\vect{y}}$ (dotted lines) with $\eta=+1$. 
The case $\eta=-1$ yields a similar distribution, shifted slightly along $\vect{q}$.
For $\vect{q}$ along $x$, the sum in Eq.~\eqref{Eq:Jz} receives a larger contribution from $\alpha$-magnon $\vect{k}$-states (red lines) carrying spin $+\hbar$ and velocity $v_x^\alpha>0$ than from the corresponding $\beta$-magnon states (blue lines), resulting in a positive spin current along $x$. 
For $\vect{q}$ along $y$, the dominant contribution arises from $\beta$-magnons (blue lines) with spin $-\hbar$ and velocity $v_y^\beta>0$, producing a negative spin current along $y$. 
For $\omega_0^{(2)} = 2.0\omega_0$, the resonant processes are two-magnon processes, with the corresponding dominant $\vect{k}$ states shown in the inset of Fig.~\ref{fig:j=2SpinCurrent}(b). 
In this regime, $\beta$-magnons with $v_x^\beta>0$ dominate for $\vect{q}$ along $x$, resulting in a negative spin current along this direction, while the situation is reversed for $\vect{q}$ along $y$. 
Since the resonant $\alpha$ and $\beta$ states are only weakly shifted in momentum space for the two-magnon processes compared with the single-magnon case [inset of \ref{fig:j=2SpinCurrent}(b) vs.  \ref{fig:j=2SpinCurrent}(a)], the spin current is smaller.

The picture becomes more intricate when $\omega_0^{(2)}$ approaches $\omega_0$, corresponding to the lower bound of $f_{2\gamma}$. 
In this regime, the resonant magnon states satisfy either $f_{1\gamma}(\vect{k},\vect{q},\eta)=\tilde{\omega}_2(\vect{q})$  or  $f_{2\gamma}(\vect{k},\vect{q},\eta)=\tilde{\omega}_2(\vect{q})$,  depending on the magnitude of $|\vect{q}|$. 
This is illustrated in Fig.~\ref{fig:j=2SpinCurrent}(d), which shows $f_{1\gamma}$ and $f_{2\gamma}$ as functions of $ak_x$ for $a\vect{q}= 0.1 \pi/2 \hat{\vect{x}}$ (solid lines) and $a\vect{q}= 0.5 \pi/2 \hat{\vect{x}}$ (dotted lines), together with the phonon energy $\tilde{\omega}_2(\vect{q})$ for $\omega_0^{(2)}=1.3\omega_0$.  As $|\vect{q}|$ increases for a fixed direction of $\vect{q}$, the resonant processes switch from two-magnon to single-magnon character, as shown in Fig.~\ref{fig:j=2SpinCurrent}(c). 
Consequently, reversing $J_i^z(\vect{q})$ does not require a full inversion of $\vect{q}$; a modest change in its magnitude, and hence in the phonon frequency, is sufficient.
A reversal of $J_i^z(\vect{q})$ is also expected at fixed $\vect{q}$ if two phonon bands, $\tilde{\omega}_2$ and $\tilde{\omega}_3$, have characteristic frequencies in regimes dominated by single- and two-magnon processes, respectively [e.g., $\omega_0^{(2)}=0.1\omega_0$ and $\omega_0^{(3)}=2.0\omega_0$]. Selective excitation of $\tilde{\omega}_2(\vect{q})$ and $\tilde{\omega}_3(\vect{q})$ would then generate magnon spin currents propagating in opposite directions.

%%%%%%%%%%%%%%%%%%%%%%%%%%%%%%%%%%%%%%%%%%%%%%%%%%%%%%%%%%%%%%%%%%%%%%%%%%%%%%
%Summary & Acknowledgements:
%%%%%%%%%%%%%%%%%%%%%%%%%%%%%%%%%%%%%%%%%%%%%%%%%%%%%%%%%%%%%%%%%%%%%%%%%%%%%
Both the $d$-wave symmetry of the phonon-induced spin current and its sign reversal upon tuning the excitation frequency constitute symmetry-governed fingerprints of altermagnetism in magnon transport. Both effects originate from the momentum-dependent chiral magnon-band splitting and vanish in the limit $\Delta J_2\rightarrow 0$. They are therefore generic features that result from the symmetry of altermagnets: the spin-dependent resonant $\vect{k}$ states entering Eq.~\eqref{eq:DistributionFunctions} are determined by the symmetry of the chiral band splitting, which is intrinsic to altermagnetic order. Furthermore, a detailed analysis of the quasi-two-dimensional oxychalcogenides KV$_2$Se$_2$O~\cite{Jiang2025} and Rb$_{1-\delta}$V$_2$Te$_2$O~\cite{Zhang2025}, as well as the insulating $d$-wave altermagnet La$_2$O$_3$Mn$_2$Se$_2$~\cite{Wei2025}, has shown that the Lieb-lattice description captures essential aspects of the spin physics. Specifically, the Lieb-lattice geometry is experimentally realized in the V\(_2\)O surface plane of KV\(_2\)Se\(_2\)O~\cite{Yang:arxiv}, highlighting the relevance of our work also for real materials.

A possible route toward coherent phonon excitation at finite momentum is provided by hard-x-ray transient-grating techniques. Recent experiments have demonstrated coherent longitudinal acoustic phonons in SrTiO$_3$ at momenta corresponding to $\sim 10\%$ of the distance from $\Gamma$ to the zone boundary~\cite{Li2026}. The available x-ray photon momentum is not restricted to this excitation range, as much higher energy x-rays are available at x-ray sources. This shows that substantially larger phonon momenta are experimentally accessible, allowing one to experimentally probe our theoretical finding.

To quantitatively estimate the magnitude of the magnon spin current, we convert the two-dimensional current in Fig.~\ref{fig:j=2SpinCurrent} to a three-dimensional current density by assuming an effective layer thickness $t_{\mathrm{eff}}=2a=8$~\AA. This yields
$J_i^z\sim 0.01-0.1 (J_1/4\pi^2 a t_{\mathrm{eff}} )\sim 3\times 10^{-6}-3\times 10^{-5}$~J/m$^2$
corresponding to a strong and experimentally accessible spin-current density. This estimate applies to a strongly driven nonequilibrium magnon population generated by coherent phonons with an amplitude of approximately $1\%$ of the lattice spacing $a$. Since the induced spin current is quadratic in the phonon amplitude, reducing the amplitude by one order of magnitude lowers the current by two orders of magnitude, while still leaving it within an experimentally relevant range.
Our results thus establish altermagnets as a promising platform for tunable magnon spin transport, with potential applications in magnon transistors and magnon-current-driven magnetic switching. An interesting direction for future work is also the inverse effect, in which selectively driven magnons generate phononic orbital currents, a phenomenon which have recently raised significant interest.

KMDH acknowledges funding from the Research Council of Norway via Project No. 334202.
MK acknowledges support from the Research Council of Norway through its Centers of Excellence funding scheme, Project No. 262633 ”QuSpin”. \newpage

%%%%Supplemental Material%%%%%%%%%%%%%%%%%%%%%%%%%%%%%%%%%%%%%%%%%%%%%%%%%%%%
\begin{widetext}

\section{Supplemental Material \\
Tunable Phonon-Driven Magnon Spin Currents in Altermagnets}

%%%%%%% Phonon model %%%%%%
\section{Phonon system}
The position of atom $k$ in unit cell $l$ is denoted by, $\vect{R} \left(\begin{smallmatrix} l \\ k \end{smallmatrix}\right)=\vect{r}\left(\begin{smallmatrix} l \\ k \end{smallmatrix}\right)+\vect{u}\left(\begin{smallmatrix} l \\ k \end{smallmatrix}\right)$, where $\vect{r}$ is the equilibrium position and $\vect{u}$ is the displacement vector. The potential energy $\Phi$ of the system to second order in the displacement $\vect{u}$ is
\begin{equation}
\label{potential energy}
    \Phi = \Phi_0 + \frac{1}{2}\sum_{\substack{l,k,\alpha \\ l',k',\beta}} \Phi_{\alpha\beta} \left(\begin{smallmatrix} l & l' \\ k & k' \end{smallmatrix}\right) u_\alpha\left(\begin{smallmatrix} l \\ k \end{smallmatrix}\right) u_\beta \left(\begin{smallmatrix} l' \\ k' \end{smallmatrix}\right)
\end{equation}
where the force constants $\Phi_{\alpha\beta}\left(\begin{smallmatrix} l & l' \\ k & k' \end{smallmatrix}\right)$ are defined as
\begin{equation}
\begin{split}
    \Phi_{\alpha\beta}
    \left(\begin{smallmatrix}
    l & l' \\
    k & k'
    \end{smallmatrix}\right)  = \left[\frac{\partial \Phi}{\partial u_\alpha \left(\begin{smallmatrix}
    l \\
    k
    \end{smallmatrix}\right)\partial u_\beta \left(\begin{smallmatrix}
    l' \\
    k'
    \end{smallmatrix}\right)}\right]_0 \text{,}  \left(\begin{smallmatrix}
    l \\
    k
    \end{smallmatrix}\right) \neq \left(\begin{smallmatrix}
    l' \\
    k'
    \end{smallmatrix}\right) 
\end{split}
\end{equation}
and obeys
\begin{equation}
\label{L=0 force constants}
    \Phi_{\alpha\beta}
   \left(\begin{smallmatrix}
    l & l \\
    k & k
    \end{smallmatrix}\right)  = -\sum_{\left(\begin{smallmatrix}
    l' \\
    k'
    \end{smallmatrix}\right) \neq \left(\begin{smallmatrix}
    l \\
    k
    \end{smallmatrix}\right)}
\Phi_{\alpha\beta}\left(\begin{smallmatrix}
    l & l' \\
    k & k'
    \end{smallmatrix}\right)
\text{\quad and \quad}
    \Phi_{\alpha\beta}\left(\begin{smallmatrix}
    l & l' \\
    k & k'
    \end{smallmatrix}\right)  = \Phi_{\beta\alpha}\left(\begin{smallmatrix}
    l' & l \\
    k' & k
    \end{smallmatrix}\right).
\end{equation} 
The force on atom $\left(\begin{smallmatrix} l \\ k \end{smallmatrix}\right)$ in the $\alpha$-direction is hence
\begin{equation}
\begin{split}
    F_\alpha\left(\begin{smallmatrix} l \\ k \end{smallmatrix}\right) = -\sum_{l',k',\beta}\Phi_{\alpha\beta}\left(\begin{smallmatrix}
    l & l' \\
    k & k'
    \end{smallmatrix}\right)u_\beta \left(\begin{smallmatrix} l' \\ k' \end{smallmatrix}\right),
\end{split}
\end{equation}
resulting in the equations of motion
\begin{equation}
\label{EOM}
    m_k\Ddot{u}_\alpha\left(\begin{smallmatrix} l \\ k \end{smallmatrix}\right) =-\sum_{l',k',\beta}\Phi_{\alpha\beta}\left(\begin{smallmatrix}
    l & l' \\
    k & k'
    \end{smallmatrix}\right)u_\beta \left(\begin{smallmatrix} l' \\ k' \end{smallmatrix}\right).
\end{equation}
The displacement vector in the $\alpha$-direction of atom $\left(\begin{smallmatrix} l \\ k \end{smallmatrix}\right)$ in phonon mode $(n,\vect{q})$ can be written as~\cite{Bruesch1982Phonons}
\begin{equation}
\label{displacement_vector}
    u_\alpha\left(\begin{smallmatrix} l \\ k \end{smallmatrix}\right)=\Re{\frac{1}{\sqrt{Nm_k}}A_n(\vect{q})e_{n,\alpha}(k|\vect{q})e^{\left[i\left(\vect{q}\cdot \vect{r}\left(\begin{smallmatrix} l \\ k \end{smallmatrix}\right)-\tilde{\omega}_n(\vect{q})t\right)\right]}}.
\end{equation}
Here, $n$ is the eigenmode with the corresponding eigenvector $\vect{e}_n(k|\vect{q})$ and eigenvalue $\tilde{\omega}_n^2(\vect{q})$ of the dynamical matrix, $N$ is the number of unit cells, $m_k$ is the mass of the atom, $A_n(\vect{q})$ parametrizes the amplitude, and $\vect{q}$ denotes the wavevector. Inserting this expression into the equations of motion (\ref{EOM}) gives the eigenvalue problem in Eq. (1) and the entries of the dynamical matrix in Eq. (3) of the main manuscript. The potential energy in Eq.~(\ref{potential energy}) is invariant under a space-group operation $g=[\vect{S}|\vect{v}+\vect{t}]$, defined through
\begin{equation}
    g\vect{R}\left(\begin{smallmatrix} l \\ k \end{smallmatrix}\right) = g\vect{r}\left(\begin{smallmatrix} l \\ k \end{smallmatrix}\right)+\vect{S}\vect{u}\left(\begin{smallmatrix} l \\ k \end{smallmatrix}\right) \text{, \;\;} g\vect{r}\left(\begin{smallmatrix} l \\ k \end{smallmatrix}\right) = \vect{S}\vect{r}\left(\begin{smallmatrix} l \\ k \end{smallmatrix}\right)+\vect{v}+\vect{t} \equiv \vect{r}\left(\begin{smallmatrix} gl \\ gk \end{smallmatrix}\right) .
\end{equation}
This invariance implies $\vect{\Phi}\left(\vect{r}\left(\begin{smallmatrix} l \\ k \end{smallmatrix}\right)+\vect{u}\left(\begin{smallmatrix} l \\ k \end{smallmatrix}\right)\right) = \vect{\Phi}\left(\vect{r}\left(\begin{smallmatrix} gl \\ gk \end{smallmatrix}\right)+\vect{S}\vect{u}\left(\begin{smallmatrix} l \\ k \end{smallmatrix}\right)\right)$,
which leads directly to the symmetry relation in Eq.~(4) of the main text. 
Here, $\vect{S}\in\mathrm{O}(3)$ is the point-group operation, $\vect{v}$ is associated with glide and screw symmetries, and $\vect{t}$ denotes a lattice translation. 
Restricting to symmorphic point group operations of the form $g=[\vect{S}|0]$, the tensorial form of the force matrices $\vect{\Phi}(\vect{L})$ in our problem can be determined from the symmetry operations of $C_{4v}$. 
For the different bonds, we obtain the following expressions
\begin{equation}
\label{force_constants}
\begin{split}
    &\vect{\Phi}(\vect{L}_1)  =\vect{S}_{\sigma_x}\cdot \vect{\Phi}(\vect{L}_1)\cdot \vect{S}_{\sigma_x}^T= -\begin{bmatrix}
        k_1 & 0 & k_4 \\
        0 & k_2 & 0 \\
        k_4' & 0 & k_3
    \end{bmatrix} \equiv \vect{\Phi}_1 \text{, \;\;} \vect{\Phi}(\vect{L}_2) =\vect{S}_{C^+_{4z}}\cdot \vect{\Phi}(\vect{L}_1)\cdot \vect{S}_{C_{4z}^+}^T= -\begin{bmatrix}
        k_2 & 0 & 0 \\
        0 & k_1 & k_4 \\
        0 & k_4' & k_3
    \end{bmatrix} \equiv \vect{\Phi}_2 \text{, \;\;}\\
    & \vect{\Phi}(\vect{L}_3)  =\vect{S}_{C_{2z}}\cdot \vect{\Phi}(\vect{L}_1)\cdot \vect{S}_{C_{2z}}^T= -\begin{bmatrix}
        k_1 & 0 & -k_4 \\
        0 & k_2 & 0 \\
        -k_4' & 0 & k_3
    \end{bmatrix} \equiv \vect{\Phi}_3  \text{, \;\;} \vect{\Phi}(\vect{L}_4)  =\vect{S}_{C_{4z}^-}\cdot \vect{\Phi}(\vect{L}_1)\cdot \vect{S}_{C_{4z}^-}^T= -\begin{bmatrix}
        k_2 & 0 & 0 \\
        0 & k_1 & -k_4 \\
        0 & -k_4' & k_3
    \end{bmatrix} \equiv \vect{\Phi}_4,\\
\end{split}
\end{equation}
where $k_1,k_2, k_3,k_4,k_4'$ are phenomenological parameters. The parameters $k_4$ and $k_4'$ result from broken inversion and vanish in the case of $D_{4h}$ symmetry. The force constants $\vect{\Phi}\left(\begin{smallmatrix} l & l \\ k & k \end{smallmatrix}\right) \equiv \vect{\Phi}(\vect{L}=0)$ for the non-magnetic atom $\vect{\Phi}_n$, spin up $\vect{\Phi}_\uparrow$ and spin down $\vect{\Phi}_\downarrow$ atom can be found from the first relation in Eq. (\ref{L=0 force constants}): 
\begin{equation}
\begin{split}
\label{L=0 force_constants}
    \vect{\Phi}_n \equiv \vect{\Phi}(\vect{L}=\vect{0}) & =\begin{bmatrix}
        2(k_1+k_2) & 0 & 0 \\
        0 & 2(k_1+k_2) & 0 \\
        0 & 0 & 4k_3
    \end{bmatrix} \text{, \;\;}
     \vect{\Phi}_\uparrow \equiv \vect{\Phi}(\vect{L}=\vect{0}) =\begin{bmatrix}
        2k_1 & 0 & 0 \\
        0 & 2k_2 & 0 \\
        0 & 0 & 2k_3
    \end{bmatrix}, \\
    \vect{\Phi}_\downarrow \equiv \vect{\Phi}(\vect{L}=\vect{0}) & =\begin{bmatrix}
        2k_2 & 0 & 0 \\
        0 & 2k_1 & 0 \\
        0 & 0 & 2k_3
    \end{bmatrix} 
\end{split}
\end{equation}
Eqs. (\ref{force_constants}, \ref{L=0 force_constants}) can then be used in Eq. (3) of the main manuscript to construct the entries of the dynamical matrix,
\begin{equation}
\begin{split}
    \mathbf{D}_{11} & = \frac{1}{m_s}\vect{\Phi}_\uparrow \text{, \;\;}
    \mathbf{D}_{13} = \frac{1} {\sqrt{m_sm_n}}[\vect{\Phi}_3^Te^{iaq_x}+\vect{\Phi}_1^Te^{-iaq_x}] \text{, \;\;}
    \mathbf{D}_{23}  = \frac{1}{\sqrt{m_sm_n}}[\vect{\Phi}_4^Te^{iaq_y}+\vect{\Phi}_2^Te^{-iaq_y}] \text{, \;\;}
    \mathbf{D}_{22}  = \frac{1}{m_s}\vect{\Phi}_\downarrow \\
    \mathbf{D}_{33} & = \frac{1}{m_n}\vect{\Phi}_n \text{, \;\;}
    \mathbf{D}_{31} = \frac{1}{\sqrt{m_sm_n}}[\vect{\Phi}_1e^{iaq_x}+\vect{\Phi}_3e^{-iaq_x}] \text{, \;\;}
    \mathbf{D}_{32} = \frac{1}{\sqrt{m_sm_n}}[\vect{\Phi}_2e^{iaq_y}+\vect{\Phi}_4e^{-iaq_y}], 
\end{split}
\end{equation}
where $a$ is the lattice spacing and $m_n=m_3$, $m_s=m_1=m_2$ are the masses of the non-magnetic atom and the spin atoms, respectively. 
We solve the eigenvalue problem in Eq.~(1) of the main text perturbatively in the vicinity of the $\Gamma$ point by expanding the dynamical matrix to second order in $\vect{q}$. 
The system have six optical modes, which for $m_n/m_s=2$ are characterized by the following energy bands
\begin{equation}
\begin{split}
    &\tilde{\omega}_1(\vect{q}) = \sqrt{\frac{2k_3}{m_s}} \text{, \;\;} \tilde{\omega}_2(\vect{q})=\sqrt{\frac{k_3}{2m_s}\left[8-a^2\left(q_x^2+q_y^2\right)\right] } \\
    &\tilde{\omega}_{3a}(\vect{q}) = \sqrt{ \frac{1}{2m_s}\left[3k_1+3k_2-\gamma_1\right]-\frac{a^2}{4m_s \gamma_1}\left[k_2\gamma_1q_x^2-5k_2^2q_x^2+k_1\gamma_1q_y^2-5k_1^2q_y^2+3k_1k_2(q_x^2+q_y^2)\right] } \\
    &\tilde{\omega}_{3b}(\vect{q}) = \sqrt{ \frac{1}{2m_s }\left[3k_1+3k_2-\gamma_1\right]-\frac{a^2}{4m_s \gamma_1} \left[k_1\gamma_1q_x^2-5k_1^2q_x^2+k_2\gamma_1q_y^2-5k_2^2q_y^2+3k_1k_2(q_x^2+q_y^2)\right] } \\
    &\tilde{\omega}_{4a}(\vect{q}) = \sqrt{ \frac{1}{2m_s}\left[3k_1+3k_2+\gamma_1\right]- \frac{a^2 \left(k_1+k_2+\gamma_1\right) }{2m_s\gamma_1(3k_2-k_1+\gamma_1)}\left[ 3k_2(k_2-k_1)q_x^2+k_2\gamma_1q_x^2+2k_1^2q_y^2  \right] } \\
    &\tilde{\omega}_{4b}(\vect{q}) = \sqrt{ \frac{1}{2m_s}\left[3k_1+3k_2+\gamma_1\right]-\frac{a^2\left(k_1+k_2+\gamma_1\right)}{2m_s\gamma_1(3k_1-k_2+\gamma_1)}\left[  3k_1(k_1-k_2)q_x^2+k_1\gamma_1q_x^2+2k_2^2q_y^2 \right] } \\
\end{split}
\end{equation}
where
\begin{equation}
    \gamma_1 = \sqrt{9k_1^2-14k_1k_2+9k_2^2}.
\end{equation}
Here, $n=1,2$ denote two nondegenerate bands, whereas the bands $n=3,4$ are degenerate at $\vect{q}=0$. 
Away from the $\Gamma$ point, these split into the modes $3a,3b$ and $4a,4b$. 
In the analysis of tunable phonon-driven magnon spin currents, we focus on the mode $n=2$. 
A comparison between the perturbative solution of $\tilde{\omega}_2(\vect{q})$ and an exact numerical solution is provided in Fig.~\ref{Sfig:1} 
The corresponding eigenvector (normalized to order $\vect{q}$) is given by
\begin{equation}
\label{eigenvectors}
    \vect{e}_{2}(\textbf{q}) = \left[
        ia_x^{(2)}q_x, \;
        ia_y^{(2)} q_y, \;
        \sin{\theta^{(2)}}\cos{\phi^{(2)}}, \;
        ib_x^{(2)}q_x, \;
        ib_y^{(2)}q_y, \;
        \sin{\theta^{(2)}}\sin{\phi^{(2)}}, \;
        ic_x^{(2)}q_x, \;
        ic_y^{(2)}q_y, \;
        \cos {\theta^{(2)}} 
    \right]^T 
\end{equation}
where $\theta^{(2)}= -\pi/4$, $\phi^{(2)}= \pi/4$. The parameters $a_\alpha^{(2)}$, $b_\alpha^{(2)}$, and $c_\alpha^{(2)}$ with $\alpha\in \{ x,y \}$ are
\begin{equation}
\begin{aligned}
a_x^{(2)} &= \frac{a \left[2 k_3 (-3 k_2 + 4 k_3) k_{4}^{'} + k_1 (k_2 - 2 k_3)(k_4 + k_{4}^{'})\right]}
{8 k_3 \left(2 k_1 k_2 - 3 (k_1 + k_2) k_3 + 4 k_3^2\right)}, \\
b_x^{(2)} &= \frac{a k_2 \left[-2 k_3 k_4 + k_1 (k_4 + k_{4}^{'})\right]}
{8 k_3 \left(2 k_1 k_2 - 3 (k_1 + k_2) k_3 + 4 k_3^2\right)}, \\
c_x^{(2)} &= \frac{a (k_2 - 2 k_3)\left[-2 k_3 k_4 + k_1 (k_4 + k_{4}^{'})\right]}
{4 \sqrt{2}\, k_3 \left(2 k_1 k_2 - 3 (k_1 + k_2) k_3 + 4 k_3^2\right)},
\end{aligned}
\quad
\begin{aligned}
a_y^{(2)} &= \frac{a k_2 \left[-2 k_3 k_4 + k_1 (k_4 + k_{4}^{'})\right]}
{8 k_3 \left(2 k_1 k_2 - 3 (k_1 + k_2) k_3 + 4 k_3^2\right)}, \\
b_y^{(2)} &= \frac{a \left[2 k_3 (-3 k_2 + 4 k_3) k_{4}^{'} + k_1 (k_2 - 2 k_3)(k_4 + k_{4}^{'})\right]}
{8 k_3 \left(2 k_1 k_2 - 3 (k_1 + k_2) k_3 + 4 k_3^2\right)}, \\
c_y^{(2)} &= \frac{a (k_2 - 2 k_3)\left[-2 k_3 k_4 + k_1 (k_4 + k_{4}^{'})\right]}
{4 \sqrt{2}\, k_3 \left(2 k_1 k_2 - 3 (k_1 + k_2) k_3 + 4 k_3^2\right)}.
\end{aligned}
\end{equation}
The parameters $a_\alpha^{(2)}$, $b_\alpha^{(2)}$, and $c_\alpha^{(2)}$ induce circular atomic motion. 
Notably, they vanish in the limit $k_4,k_4' \to 0$, indicating that they originate from the broken inversion symmetry of the system.
From the displacement vector in Eq. (\ref{displacement_vector}), we find the dc phonon angular momentum (here, $T_{n}(\vect{q})=2\pi/\tilde{\omega}_n (\vect{q})$)
\begin{equation}
    \vect{J}_n(\vect{q})= \sum_{lk}\frac{1}{T_{n}(\vect{q}) } \int_{0}^{T_{n}(\vect{q}) }  \vect{u} \left(\begin{smallmatrix} l \\ k \end{smallmatrix}\right) \times m_k \dot{\vect{u}}\left(\begin{smallmatrix} l \\ k \end{smallmatrix}\right) {\rm dt} =\frac{1}{2}\abs{A_n(\vect{q})}^2\tilde{\omega}_n(\vect{q})\sum_{k=1,2,3}\Im{\vect{e}_{n}(k|\vect{q})^* \times\vect{e}_{n}(k|\vect{q})}.
\end{equation}
Substituting the eigenvector for $n=2$ yields the following expression to order $\vect{q}$ (in the expansion for the eigenvector)
\begin{equation}
\vect{J}_{2}(\vect{q}) =\frac{1}{2}\abs{A_2(\vect{q})}^2\tilde{\omega}_2(\vect{q})\left( a_x^{(2)}+ a_y^{(2)}- \sqrt{2} c_x^{(2)} \right)  \vect{q}\times\hat{\vect{z}}. 
\end{equation}
Thus, the angular momentum lies in the $xy$ plane, perpendicular to the propagation direction $\vect{q}$.
The $\mathcal{O}(\vect{q}^2)$ corrections to the eigenvector yield only real expansion coefficients and thus merely renormalize the linear polarization arising from the leading $\mathcal{O}(1)$ terms. 
They are therefore omitted in Eq.~\eqref{eigenvectors}.
\begin{figure}[ht] 
\centering 
\includegraphics[width=1\linewidth]{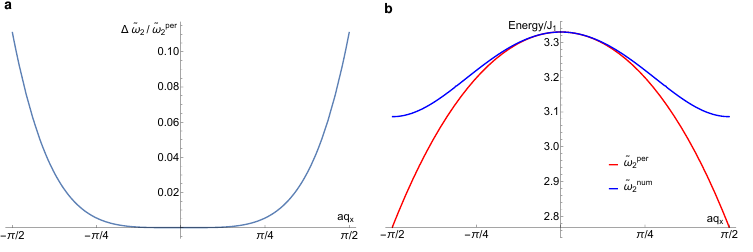}  
\caption{ Comparison between the perturbative and exact numerical solutions for $\tilde{\omega}_2(\vect{q})$.
(\textbf{a}) Relative error of the perturbative solution.
(\textbf{b}) Perturbative (red) and numerical (blue) results for $\tilde{\omega}_2(\vect{q})$. Here, we have used the parameter values:  $k_1/m_s = 0.2 \omega_0^2$, $k_2/m_s = 0.6 \omega_0^2$, $k_3/m_s = 0.42 \omega_0^2$, $k_4/m_s= k_4^{'}/m_s = 0.1 \omega_0^2$ where 
$\omega_0=4S\sqrt{K_z^2-4J_1K_z}$ with $J_1= 25.7$~meV, $K_z/J_1=-0.1$, and $S=1$. }
\label{Sfig:1}
\end{figure} 

%%%%%%% Magnon model %%%%%%
\section{Magnon system}
Starting from Eq.~(7) of the main text, we write the spin operators as
\begin{equation}
\label{ladder operators}
\begin{split}
     S^A_{i,x} & = \frac{1}{2}\big(S^A_{i,+}+S^A_{i,-}\big)\text{, \;\;} S^A_{i,y}=\frac{1}{2i}\big(S^A_{i,+}-S^A_{i,-}\big) \text{, \;\;}
     S^B_{i,x} = \frac{1}{2}\big(S^B_{i,+}+S^B_{i,-}\big)\text{, \;\;} S^B_{i,y}=\frac{1}{2i}\big(S^B_{i,+}-S^B_{i,-}\big),
\end{split}
\end{equation}
and represent them in terms of Holstein–Primakoff bosons via the expressions~\cite{Auerbach}
\begin{equation}
\label{HP transformation}
\begin{split}
    &S^A_{i,+}  =\sqrt{2S}a_i \text{, \;\;} S^A_{i,-}  =\sqrt{2S}a^\dag_i \text{, \;\;} S^A_{i,z}  =S-a^\dag_i a_i \text{, \;\;} S^B_{i,+}=\sqrt{2S}b^\dag_i \text{, \;\;} S^B_{i,-}=\sqrt{2S}b_i \text{, \;\;} S^B_{i,z}=b^\dag_i b_i-S,
\end{split}
\end{equation}
which are valid to second order in the ladder operators.
We then Fourier transform according to $a_i = \frac{1}{\sqrt{N}} \sum_\vect{k} a_\vect{k} e^{i\vect{k}\cdot\vect{r}_i}$ and $b_i = \frac{1}{\sqrt{N}} \sum_\vect{k} b_\vect{k} e^{i\vect{k}\cdot\vect{r}_i}$, where $N$ is the number of unit cells and $\vect{k}$ lies in the first Brillouin zone. 
This results in the magnon Hamiltonian in Eq. (8) of the main manuscript, where we have \begin{equation}
\begin{split}
    A_\vect{k} = & S[4J_1-2K_z+2(J_2\cos{(2k_xa)}+J_2'\cos{(2k_ya)})-2(J_2+J_2')] , \\
    C_\vect{k}  = & S[4J_1-2K_z+2(J_2'\cos{(2k_xa)}+J_2\cos{(2k_ya)})-2(J_2+J_2')] , \\
    B_\vect{k}  = & 2SJ_1[\cos{(k_xa+k_ya)}+\cos{(k_xa-k_ya)}] .
\end{split}
\end{equation}
The magnon Hamiltonian is diagonalized via a bosonic Bogoliubov–de Gennes (BdG) transformation~\cite{Auerbach}. 
Requiring the off-diagonal terms to vanish yields the condition
$[A_\vect{k}+C_\vect{k}]u_\vect{k}^*v_\vect{k}+B_\vect{k}[v_\vect{k}^2+(u_\vect{k}^*)^2]=0$,
which must be solved together with the constraint
$\abs{u_\vect{k}}^2-\abs{v_\vect{k}}^2=1$
imposed by the bosonic commutation relations. 
A possible solution is 
\begin{equation}
\label{u_k,v_k}
\begin{split}
u_\vect{k} &= \frac{i}{\sqrt{2}}
\sqrt{\frac{A_\vect{k}+C_\vect{k}}
{\sqrt{(A_\vect{k}+C_\vect{k})^2-4B_\vect{k}^2}} + 1}, \\
v_\vect{k} &= \frac{i}{\sqrt{2}}
\sqrt{\frac{A_\vect{k}+C_\vect{k}}
{\sqrt{(A_\vect{k}+C_\vect{k})^2-4B_\vect{k}^2}} - 1},
\end{split}
\end{equation}
which diagonalizes the Hamiltonian with magnon energies 
$\omega_\alpha(\vect{k})$ and $\omega_\beta(\vect{k})$ given in the main text.

%%%%%%% Magnon-phonon coupling %%%%%%
\section{Magnon-phonon coupling}
\subsection{Interaction Hamiltonian}
The magnon–phonon coupling is obtained by expanding the exchange interactions to linear order in the atomic displacements:
$H_{I1} = \alpha_1 \sum_{\langle i,j \rangle} \hat{\vect{n}}_{ij}\!\cdot\!\Delta\vect{u}_{ij}\, \vect{S}_i^A \!\cdot\! \vect{S}_j^B$, 
$H_{I2} = \sum_{\langle i_x,j_x \rangle} \hat{\vect{n}}_{ij}\!\cdot\!\Delta\vect{u}_{ij} [ \alpha_2 \vect{S}_i^A \!\cdot\! \vect{S}_j^A + \alpha_2' \vect{S}_i^B \!\cdot\! \vect{S}_j^B ]$, and 
$H_{I3} = \sum_{\langle i_y,j_y \rangle} \hat{\vect{n}}_{ij}\!\cdot\!\Delta\vect{u}_{ij} [ \alpha_2 \vect{S}_i^B \!\cdot\! \vect{S}_j^B + \alpha_2' \vect{S}_i^A \!\cdot\! \vect{S}_j^A ]$, 
where $\alpha_1=\partial J_1/\partial a$, $\alpha_2=\partial J_2/\partial a$, and $\alpha_2'=\partial J_2'/\partial a$ for lattice constant $a$. 
Here, $\Delta\vect{u}_{ij}=\vect{u}_j-\vect{u}_i$ denotes the relative displacement, and $\hat{\vect{n}}_{ij}$ is the unit vector pointing from site $i$ to $j$. 
Expressing the spin operators in terms of Holstein–Primakoff bosons and transforming to momentum space yields
\begin{equation}
\label{H_I 1}
\begin{split}
   H_I = 2i\hbar^2 S \sum_{\substack{n,\vect{k},\vect{q} \\ i\in\{1,2\}}}&\Big\{\alpha_1\left[\hat{\vect{e}}_{i}\cdot\vect{u}_2(n,\vect{q})a^\dag_{\vect{k}+\vect{q}}a_{\vect{k}}+\hat{\vect{e}}_{i}\cdot\vect{u}_1(n,\vect{q})b^\dag_{\vect{k}+\vect{q}} b_{\vect{k}}\right]\sin{\left[\vect{q}\cdot\vect{\vect{e}}_i\right]} \\
   &+\alpha_1\left[\hat{\vect{e}}_{i}\cdot\vect{u}_1(n,\vect{q})b^\dag_{\vect{q}-\vect{k}}a^\dag_{\vect{k}} -\hat{\vect{e}}_{i}\cdot\vect{u}_1(n,-\vect{q})a_{\vect{k}} b_{\vect{q}-\vect{k}}\right]\sin{\left[\left(\vect{q}-\vect{k}\right)\cdot\vect{\vect{e}}_i\right]} \\
    &+\alpha_1\left[\hat{\vect{e}}_{i}\cdot\vect{u}_2(n,\vect{q})a^\dag_{\vect{q}-\vect{k}}b^\dag_{\vect{k}} -\hat{\vect{e}}_{i}\cdot\vect{u}_2(n,-\vect{q})b_{\vect{k}}a_{\vect{q}-\vect{k}}\right]\sin{\left[\left(\vect{q}-\vect{k}\right)\cdot\vect{\vect{e}}_i\right]} \\
     &+ \left[\Bar{\alpha}_2-(-1)^i  \Delta \alpha_2\right]\left[\hat{\vect{r}}_i \cdot \vect{u}_1(n,\vect{q})\right]f(\vect{r}_i, \vect{q}, \vect{k})a^\dag_{\vect{k}+\vect{q}}a_{\vect{k}} \\
    &+\left[\Bar{\alpha}_2+(-1)^i \Delta \alpha_2\right]\left[\hat{\vect{r}}_i \cdot \vect{u}_2(n,\vect{q})\right]f(\vect{r}_i, \vect{q}, \vect{k})b^\dag_{\vect{k}+\vect{q}}b_{\vect{k}} \Big\}
\end{split}
\end{equation}
where $\vect{e}_1=(a,a,0), \vect{e}_2=(-a,a,0), \vect{r}_1=(2a,0,0), \vect{r}_2 = (0,2a,0)$ for lattice spacing $a$, $f(\vect{r}_i, \vect{q}, \vect{k})=\sin{[\vect{k}\cdot\vect{r}_i]}+\sin{[\vect{q}\cdot\vect{r}_i]}-\sin{[(\vect{k}+q)\cdot\vect{r}_i]}$,
$\Bar{\alpha}_2= (\alpha_2 + \alpha_2')/2$ and $\Delta\alpha_2= (\alpha_2 - \alpha_2')/2$. 
For the spin-up ($k=1$) and spin-down ($k=2$) sublattices, the displacement fields are expanded in phonon eigenmodes $(n,\vect{q})$ as
\begin{equation}
\vect{u}
\left(
\begin{smallmatrix}
l \\ k
\end{smallmatrix}
\right)
=
\sum_{n,\vect{q}}
\vect{u}_k(n,\vect{q})\,
e^{i\vect{q}\cdot\vect{r}
\left(
\begin{smallmatrix}
l \\ k
\end{smallmatrix}
\right)},
\qquad k=1,2.
\end{equation}
The mode amplitudes can be expressed in terms of the normal coordinates $Q_n(\vect{q})$ and phonon eigenvectors $\vect{e}_n(k|\vect{q})$ as~\cite{Bruesch1982Phonons}
\begin{equation}
\vect{u}_k(n,\vect{q})
=
\sqrt{\frac{1}{N m_k}}\,
\vect{e}_n(k|\vect{q})\,
Q_n(\vect{q}),
\qquad
Q_n^*(\vect{q}) = Q_n(-\vect{q}).
\end{equation}
Substituting this expression into Eq.~(\ref{H_I 1}) yields
\begin{equation}
\begin{split}\label{Eq:HI1}
    H_I =\sum_{n,\vect{k},\vect{q}}\big[&\Lambda_a(n,\vect{q},\vect{k})Q_n(\vect{q})a^\dag_{\vect{k}+\vect{q}}a_{\vect{k}}
    +\Lambda_b(n,-\vect{q},-\vect{k})Q^*_n(\vect{q})b^\dag_{-\vect{k}-\vect{q}}b_{-\vect{k}}\\
    &+\Lambda_0(n,-\vect{q},\vect{k})Q^*_n(\vect{q})b^\dag_{-\vect{q}-\vect{k}}a^\dag_{\vect{k}}
    +\Lambda^*_0(n,-\vect{q},\vect{k})Q_n(\vect{q})b_{-\vect{q}-\vect{k}}a_{\vect{k}}\big]
\end{split}
\end{equation}
where
\begin{equation}
\begin{split}
     \Lambda_0(n,\vect{q},\vect{k}) &= \sqrt{\frac{1}{N m_s}}\sum_{i=1,2}\big[\Gamma_0(i,\vect{q}-\vect{k})\hat{\vect{e}}_i \cdot \vect{e}_n(1|\vect{q})+\Gamma_0(i,\vect{k})\hat{\vect{e}}_i \cdot \vect{e}_n(2|\vect{q})\big], \\
     \Lambda_a(n,\vect{q},\vect{k}) &= \sqrt{\frac{1}{N m_s}}\sum_{i=1,2}\big[\Gamma_a(i,\vect{q},\vect{k})\hat{\vect{r}}_i \cdot \vect{e}_n(1|\vect{q})+\Gamma_0(i,\vect{q})\hat{\vect{e}}_i \cdot \vect{e}_n(2|\vect{q})\big], \\
     \Lambda_b(n,\vect{q},\vect{k}) &=\sqrt{\frac{1}{N m_s}}\sum_{i=1,2}\big[\Gamma_b(i,\vect{q},\vect{k})\hat{\vect{r}}_i \cdot \vect{e}_n(2|\vect{q})+\Gamma_0(i,\vect{q})\hat{\vect{e}}_i \cdot \vect{e}_n(1|\vect{q})\big], \\
     \Gamma_a(i,\vect{q},\vect{k}) &= 2i\hbar^2 S\big[\Bar{\alpha}_2-(-1)^i\Delta\alpha_2\big]f(\vect{r}_i, \vect{q}, \vect{k}), \\
     \Gamma_b(i,\vect{q},\vect{k}) & = 2i\hbar^2 S\big[\Bar{\alpha}_2+(-1)^i\Delta\alpha_2\big]f(\vect{r}_i, \vect{q}, \vect{k}), \\
      \Gamma_0(i,\vect{q}) & = 2i\hbar^2 S\alpha_1\sin{\left[\vect{e}_i\cdot\vect{q}\right]}.
\end{split}
\end{equation}
 The Holstein–Primakoff bosons are further expressed in terms of the elementary magnon operators via the BdG transformation
 $a_\vect{k}=u_\vect{k}\alpha_\vect{k}+v_\vect{k}\beta^\dag_{-\vect{k}}\text{ and } b^\dag_{-\vect{k}}=v^*_\vect{k}\alpha_\vect{k}+u^*_\vect{k}\beta^\dag_{-\vect{k}}$,
where $u_\vect{k}$ and $v_\vect{k}$ are defined in Eq.~(\ref{u_k,v_k}). 
Substituting these relations into Eq.~\eqref{Eq:HI1} yields Eq.~(10) of the main text, with the $\Omega$ prefactors given by
\begin{equation}
\begin{split}
    &\Omega_{\Bar{\alpha}\alpha}(n,\vect{q},\vect{k}) = \Lambda^a_{\Bar{\alpha}\alpha}(n,\vect{q},\vect{k}) +\Lambda^b_{\alpha\Bar{\alpha}}(n,-\vect{q},\vect{k}+\vect{q}) \text{, \;\;} \Omega_{\Bar{\beta}\beta}(n,\vect{q},\vect{k}) = \Lambda^a_{\beta \Bar{\beta}}(n,\vect{q},\vect{k}) +\Lambda^b_{\Bar{\beta}\beta}(n,-\vect{q},\vect{k}+\vect{q}) \\
    &\Omega_{\Bar{\alpha}\Bar{\beta}}(n,\vect{q},\vect{k}) = \Lambda^a_{\Bar{\alpha} \Bar{\beta}}(n,\vect{q},\vect{k}) +\Lambda^b_{\Bar{\beta}\Bar{\alpha} }(n,-\vect{q},\vect{k}+\vect{q})\text{, \;\;} \Omega_{\alpha\beta}(n,\vect{q},\vect{k}) = \Lambda^a_{\beta \alpha}(n,\vect{q},\vect{k}) +\Lambda^b_{\alpha\beta}(n,-\vect{q},\vect{k}+\vect{q})  
    %& \Omega_{\alpha \Bar{\beta}}(n,\vect{q},\vect{k}) = \Lambda^a_{\beta\Bar{\beta}}(n,\vect{q},\vect{k}) +\Lambda^b_{\alpha\Bar{\alpha}}(n,\vect{q},\vect{k})
\end{split}
\end{equation}
where
\begin{equation}
\begin{split}
    & \Lambda^a_{\Bar{\alpha}\alpha}(n,\vect{q},\vect{k}) = \Lambda_a(n,\vect{q},\vect{k})u^*_{\vect{k}+\vect{q}}u_{\vect{k}}+\Lambda^*_0(n,-\vect{q},\vect{k})v_{\vect{k}+\vect{q}}u_{\vect{k}} \\
    & \Lambda^a_{\beta \Bar{\beta}}(n,\vect{q},\vect{k}) = \Lambda_a(n,\vect{q},\vect{k})v^*_{\vect{k}+\vect{q}}v_{\vect{k}}+\Lambda^*_0(n,-\vect{q},\vect{k})u_{\vect{k}+\vect{q}}v_{\vect{k}} \\
    & \Lambda^a_{\Bar{\alpha} \Bar{\beta}}(n,\vect{q},\vect{k}) = \Lambda_a(n,\vect{q},\vect{k})u^*_{\vect{k}+\vect{q}}v_{\vect{k}}+\Lambda^*_0(n,-\vect{q},\vect{k})v_{\vect{k}+\vect{q}}v_{\vect{k}} \\
    & \Lambda^a_{\beta \alpha}(n,\vect{q},\vect{k}) = \Lambda_a(n,\vect{q},\vect{k})v^*_{\vect{k}+\vect{q}}u_{\vect{k}}+\Lambda^*_0(n,-\vect{q},\vect{k})u_{\vect{k}+\vect{q}}u_{\vect{k}} \\
    & \\
    & \Lambda^b_{\alpha\Bar{\alpha}}(n,\vect{q},\vect{k}) = \Lambda_b(n,-\vect{q},-\vect{k})v^*_{\vect{k}+\vect{q}}v_{\vect{k}}+\Lambda_0(n,-\vect{q},\vect{k})v^*_{\vect{k}+\vect{q}}u^*_{\vect{k}} \\
    & \Lambda^b_{\Bar{\beta}\beta}(n,\vect{q},\vect{k}) = \Lambda_b(n,-\vect{q},-\vect{k})u^*_{\vect{k}+\vect{q}}u_{\vect{k}}+\Lambda_0(n,-\vect{q},\vect{k})u^*_{\vect{k}+\vect{q}}v^*_{\vect{k}} \\
    & \Lambda^b_{\alpha\beta}(n,\vect{q},\vect{k}) = \Lambda_b(n,-\vect{q},-\vect{k})v^*_{\vect{k}+\vect{q}}u_{\vect{k}}+\Lambda_0(n,-\vect{q},\vect{k})v^*_{\vect{k}+\vect{q}}v^*_{\vect{k}} \\
    & \Lambda^b_{\Bar{\beta}\bar{\alpha}}(n,\vect{q},\vect{k}) = \Lambda_b(n,-\vect{q},-\vect{k})u^*_{\vect{k}+\vect{q}}v_{\vect{k}}+\Lambda_0(n,-\vect{q},\vect{k})u^*_{\vect{k}+\vect{q}}u^*_{\vect{k}}
\end{split}
\end{equation}

%%%%%%Distribution function%%%%%
\subsection{Distribution functions}
Expanding the contour-ordered Green’s function
$G_{\gamma\Bar{\gamma}}^{(\tau \tau' )}(\vect{k},t;\vect{k}',t') = -i\big\langle\mathcal{T}_c e^{-\frac{i}{{\hbar}}{\int_c}H_I(t_1)dt_1}\gamma_\vect{k}(t,\tau)\gamma^\dag_\vect{k}(t',\tau')\big\rangle_0$ to second order in $H_I$ yields
\begin{equation}
\label{eq:G alpha}
\begin{split}
    G_{\gamma\Bar{\gamma}}^{(\tau\tau')}(\vect{k},t;\vect{k}',t') & = -i \left\langle \mathcal{T}_c\gamma_\vect{k}(t,\tau)\gamma^\dag_{\vect{k}'}(t',\tau' )\right\rangle_0 + \frac{i}{2\hbar^2} \int_c dt_1 \int_c dt_2 \left\langle \mathcal{T}_c H_I(t_1)H_I(t_2)\gamma_\vect{k}(t,\tau)\gamma^{\dag}_{\vect{k}'}(t',\tau')\right\rangle_0 \\
    & =G_{\gamma\Bar{\gamma}}^{(\tau\tau')}(\vect{k},t;\vect{k}',t')_0 +\delta G_{\gamma\Bar{\gamma}}^{(\tau\tau')}(\vect{k},t;\vect{k}',t')
\end{split}
\end{equation}
where the first-order contribution vanishes. 
The first term describes the equilibrium state, while the second term captures nonequilibrium corrections induced by the magnon–phonon coupling.
We consider the system to host a single phonon mode $(n,\vect{q})$, for which the displacement field reads
\begin{equation}
\vect{u}
\left(
\begin{smallmatrix}
l \\ k
\end{smallmatrix}
\right)
=
\frac{1}{\sqrt{N m_k}}
\,\mathrm{Re}\!\left[
A_n(\vect{q})\, \vect{e}_n(k|\vect{q})\,
e^{ i\vect{q}\cdot \vect{r}
\left(
\begin{smallmatrix}
l \\ k
\end{smallmatrix}
\right)
-i\tilde{\omega}_n(\vect{q}) t}
\right].
\end{equation}
The phonon normal coordinate $Q_n(\vect{q},t)$ is treated as a classical field and, for the single-mode excitation, is given by
\begin{equation}
\begin{split}
\label{phonon normal coordinate}
Q_{n'}(\vect{q}',t)
&= \frac{1}{\sqrt{N}} \sum_{l,k} \sqrt{m_k}\,
e^{-i\vect{q}'\cdot \vect{r}
\left(
\begin{smallmatrix}
l \\ k
\end{smallmatrix}
\right)}
\, \vect{e}^*_{n'}(k|\vect{q}') \cdot 
\vect{u}
\left(
\begin{smallmatrix}
l \\ k
\end{smallmatrix}
\right) \\
&= \frac{1}{2}\delta_{\vect{q}',\vect{q}}\delta_{n',n}
A_n(\vect{q})e^{-i\tilde{\omega}_n(\vect{q})t}
+ \frac{1}{2}\delta_{\vect{q}',-\vect{q}}\delta_{n',n}
A_n^*(\vect{q})e^{i\tilde{\omega}_n(\vect{q})t}.
\end{split}
\end{equation}
This implies (here, $\eta\in \{+1,-1 \}$)
\begin{equation}
Q_n(-\eta\vect{q},t_2)\, Q_n(\eta\vect{q},t_1)
=
\frac{|A_n(\vect{q})|^2}{4}\,
e^{i\eta \tilde{\omega}_n(\vect{q})(t_2-t_1)} .
\end{equation}

We first consider excitations of the $\alpha$ magnons. 
Substituting $H_I$ from Eq.~(10) of the main text into the second term of Eq.~\eqref{eq:G alpha} and evaluating the statistical averages using Wick’s theorem~\cite{Rammer_2007}, with the phonon normal coordinate $Q_n(\vect{q},t)$ given in Eq.~\eqref{phonon normal coordinate}, yields
\begin{equation}
\begin{split}
    & \frac{i}{2\hbar^2}  \left\langle \mathcal{T}_c H_I(t_1)H_I(t_2)\alpha_\vect{k}(t,\tau)\alpha^{\dag}_{\vect{k}'}(t',\tau')\right\rangle_0 =  \frac{1}{\hbar^2} \sum_{\vect{q}_1=\pm \vect{q}} \\
    & \Omega_{\Bar{\alpha}\alpha}(n,\vect{q}_1,\vect{k})\Omega_{\Bar{\alpha}\alpha}(n,-\vect{q}_1,\vect{k}+\vect{q}_1)G_{\alpha\Bar{\alpha}}^{(\tau\tau_2)}(\vect{k};t-t_2)_0 Q_n(-\vect{q}_1,t_2)G_{\alpha\Bar{\alpha}}^{(\tau_2\tau_1)}(\vect{k}+\vect{q}_1;t_2-t_1)_0 Q_n(\vect{q}_1,t_1)G_{\alpha\Bar{\alpha}}^{(\tau_1 \tau')}(\vect{k},t_1-t')_0 + \\
    & \Omega_{\alpha\beta}(n,\vect{q}_1,\vect{k})\Omega_{\Bar{\alpha}\Bar{\beta}}(n,-\vect{q}_1,\vect{k}+\vect{q}_1)G_{\alpha\Bar{\alpha}}^{(\tau\tau_2)}(\vect{k},t-t_2)_0 Q_n(-\vect{q}_1,t_2)G_{\Bar{\beta}\beta}^{(\tau_2\tau_1)}(-\vect{k}-\vect{q}_1;t_2-t_1)_0 Q_n(\vect{q}_1,t_1)G_{\alpha\Bar{\alpha}}^{(\tau_1\tau')}(\vect{k};t_1-t')_0
\end{split}
\end{equation}
where 
$G^{(\tau\tau')}_{\alpha\bar{\alpha}}(\vect{k},t;\vect{k}',t')_0 = -i\langle \mathcal{T}_c\, \alpha_\vect{k}(t,\tau)\alpha^\dag_{\vect{k}'}(t',\tau')\rangle_0$ 
and 
$G^{(\tau\tau')}_{\bar{\beta}\beta}(\vect{k},t;\vect{k}',t')_0 = -i\langle \mathcal{T}_c\, \beta^\dag_\vect{k}(t,\tau)\beta_{\vect{k}'}(t',\tau')\rangle_0$.
Similarly, we define $G^{(\tau\tau')}_{\beta\bar{\beta}}(\vect{k},t;\vect{k}',t')_0 = -i\langle \mathcal{T}_c\, \beta_\vect{k}(t,\tau)\beta^\dag_{\vect{k}'}(t',\tau')\rangle_0$ 
and $G^{(\tau\tau')}_{\bar{\alpha}\alpha}(\vect{k},t;\vect{k}',t')_0 = -i\langle \mathcal{T}_c\, \alpha^\dag_\vect{k}(t,\tau)\alpha_{\vect{k}'}(t',\tau')\rangle_0$, which are used below.  
Mapping the Green’s functions to Keldysh space, transforming to the trigonal representation 
$\underline{G}=[G^{R},G^{K};0,G^{A}]$~\cite{Rammer_2007}, and Fourier transforming according to 
$G(t-t')=\frac{1}{2\pi}\int_{-\infty}^{\infty} e^{-i\omega(t-t')}G(\omega)\,d\omega$, 
we obtain the Green’s function of the $\alpha$ magnons,
\begin{equation}
\label{eq:Galpha}
\underline{G}_{\alpha\bar{\alpha}}(\vect{k};\omega)
=
\begin{bmatrix}
G^{R}_{\alpha\bar{\alpha}}(\vect{k};\omega) & G^{K}_{\alpha\bar{\alpha}}(\vect{k};\omega) \\
0 & G^{A}_{\alpha\bar{\alpha}}(\vect{k};\omega)
\end{bmatrix}
=
\underline{G}_{\alpha\bar{\alpha}}(\vect{k};\omega)_0
+
\underline{G}_{\alpha\bar{\alpha}}(\vect{k};\omega)_0\,
\underline{\Sigma}_{\alpha\bar{\alpha}}(\vect{k};\omega)\,
\underline{G}_{\alpha\bar{\alpha}}(\vect{k};\omega)_0,
\end{equation}
with self-energy
\begin{equation}
\begin{split}
\underline{\Sigma}_{\alpha\bar{\alpha}}(\vect{k};\omega)
= \frac{|A_n(\vect{q})|^2}{4\hbar^2}
\sum_{\eta=\pm1}
\Big[
&\Omega_{\bar{\alpha}\alpha}(n,\eta\vect{q},\vect{k})
\Omega_{\bar{\alpha}\alpha}(n,-\eta\vect{q},\vect{k}+\eta\vect{q})\,
\underline{G}_{\alpha\bar{\alpha}}
\big(\vect{k}+\eta\vect{q};\omega+\eta\tilde{\omega}_n(\vect{q})\big)_0 \\
+&\Omega_{\alpha\beta}(n,\eta\vect{q},\vect{k})
\Omega_{\bar{\alpha}\bar{\beta}}(n,-\eta\vect{q},\vect{k}+\eta\vect{q})\,
\underline{G}_{\bar{\beta}\beta}
\big(-\vect{k}-\eta\vect{q};\omega+\eta\tilde{\omega}_n(\vect{q})\big)_0
\Big].
\end{split}
\end{equation}
A corresponding calculation for the $\beta$ magnons yields
\begin{equation}
\label{eq:Gbeta}
\underline{G}_{\beta\bar{\beta}}(\vect{k};\omega)
=
\underline{G}_{\beta\bar{\beta}}(\vect{k};\omega)_0
+
\underline{G}_{\beta\bar{\beta}}(\vect{k};\omega)_0\,
\underline{\Sigma}_{\beta\bar{\beta}}(\vect{k};\omega)\,
\underline{G}_{\beta\bar{\beta}}(\vect{k};\omega)_0,
\end{equation}
with
\begin{equation}
\begin{split}
\underline{\Sigma}_{\beta\bar{\beta}}(\vect{k};\omega)
= \frac{|A_n(\vect{q})|^2}{4\hbar^2}
\sum_{\eta=\pm1}
\Big[
&\Omega_{\bar{\beta}\beta}(n,\eta\vect{q},-\vect{k}-\eta\vect{q})
\Omega_{\bar{\beta}\beta}(n,-\eta\vect{q},-\vect{k})\,
\underline{G}_{\beta\bar{\beta}}
\big(\vect{k}+\eta\vect{q};\omega+\eta \tilde{\omega}_n(\vect{q})\big)_0 \\
+&\Omega_{\alpha\beta}(n,\eta\vect{q},-\vect{k}-\eta\vect{q})
\Omega_{\bar{\alpha}\bar{\beta}}(n,-\eta\vect{q},-\vect{k})\,
\underline{G}_{\bar{\alpha}\alpha}
\big(-\vect{k}-\eta\vect{q};\omega+\eta\tilde{\omega}_n(\vect{q})\big)_0
\Big].
\end{split}
\end{equation}

The equilibrium Greens functions take the forms (here, $\gamma \in [\alpha,\beta]$)
\begin{equation}
\label{eq:EqulibriumGreens}
\begin{array}{c@{\qquad}c}
    \begin{aligned}
    G^>_{\gamma\Bar{\gamma}}(\vect{k};\omega)_0 & = -2 \pi i\left[1+n_{\gamma}^0(\vect{k})\right]\delta(\omega-\omega_\gamma(\vect{k})) &\\
    G^<_{\gamma\Bar{\gamma}}(\vect{k};\omega)_0 & = -2 \pi i n_{\gamma}^0(\vect{k}) \delta(\omega-\omega_\gamma(\vect{k})) \\
    G^K_{\gamma\Bar{\gamma}}(\vect{k};\omega)_0 & = G^>_{\gamma\Bar{\gamma}}(\vect{k};\omega)_0+G^<_{\gamma\Bar{\gamma}}(\vect{k};\omega)_0 \\
    &= -2 \pi i \coth{\left[\tilde{\beta} E_\gamma(\vect{k})/2\right]}\delta(\omega-\omega_\gamma(\vect{k})) \\
    G^R_{\gamma\Bar{\gamma}}(\vect{k};\omega)_0 & =G^A_{\gamma\Bar{\gamma}}(\vect{k};\omega)_0^*= \frac{1}{\omega-\omega_\gamma(\vect{k})+i\varepsilon},\\
    \end{aligned}
    &
    \begin{aligned}
    G^>_{\Bar{\gamma}\gamma}(\vect{k};\omega)_0 & = -2 \pi i n_{\gamma}^0(\vect{k})\delta(\omega+\omega_\gamma(\vect{k})) \\
    G^<_{\Bar{\gamma}\gamma}(\vect{k};\omega)_0 & = -2 \pi i\left[1+n_{\gamma}^0(\vect{k})\right]\delta(\omega+\omega_\gamma(\vect{k})) \\
    G^K_{\Bar{\gamma}\gamma}(\vect{k};\omega)_0 & = G^>_{\Bar{\gamma}\gamma}(\vect{k};\omega)_0+G^<_{\Bar{\gamma}\gamma}(\vect{k};\omega)_0 \\
    &= -2 \pi i \coth{\left[\tilde{\beta} E_\gamma(\vect{k})/2\right]}\delta(\omega+\omega_\gamma(\vect{k})) \\
    G^R_{\Bar{\gamma}\gamma}(\vect{k};\omega)_0 & =G^A_{\Bar{\gamma}\gamma}(\vect{k};\omega)_0^*= \frac{-1}{\omega+\omega_\gamma(\vect{k})+i\varepsilon},
    \end{aligned}
\end{array}
\end{equation}
where $\tilde{\beta}=1/k_B T$, $E_\gamma(\vect{k})=\hbar\omega_\gamma(\vect{k})$ is the magnon energy, and $n_{\gamma}^0(\vect{k}) = 1/(e^{\tilde{\beta} E_\gamma(\vect{k})}-1)$ is the Bose-Einstein distribution (throughout, we set $\hbar=1$).
The Keldysh component $G^K_{\gamma\Bar{\gamma}}(\vect{k};\omega)$ of the Green's function is determined by
\begin{equation}
\begin{split}
    G^K_{\gamma\Bar{\gamma}}(\vect{k};\omega) &= G^K_{\gamma\Bar{\gamma}}(\vect{k};\omega)_0   +\left[\underline{G}_{\gamma\Bar{\gamma}}(\vect{k};\omega)_0\underline{\Sigma}_{\gamma\Bar{\gamma}}(\vect{k};\omega)\underline{G}_{\gamma\Bar{\gamma}}(\vect{k};\omega)_0\right]_{12}
\end{split}
\end{equation}
where the subscript $12$ denotes the matrix-element. 
Equivalently, it can be obtained from the Keldysh equation~\cite{Haug_2007}  
\begin{equation}
    G^K = [1+G^R\Sigma^R]G^K_0[1+\Sigma^AG^A]+G^R\Sigma^KG^A, \label{Eq:KeldyshEq}
\end{equation}
keeping terms to second order in $H_I$. In Eq.~\eqref{Eq:KeldyshEq},
the first term encodes information about the initial state of the system at $t_0 \to -\infty$. For systems prepared in a noninteracting state at $t_0$, this term vanishes in the steady state~\cite{Haug_2007}.
Consequently, only the second term contributes to the nonequilibrium correction $\delta G^K_{\gamma\bar{\gamma}}(\vect{k};\omega)$ in steady state. 
Using the expressions in Eq.~(\ref{eq:EqulibriumGreens}), we thus obtain to second order in $H_I$
\begin{equation}
\label{eq:KeldyshComponent}
\begin{split}
    \delta G^K_{\gamma\Bar{\gamma}}(\vect{k};\omega) &= 
    G^R_{\gamma\Bar{\gamma}}(\vect{k};\omega)_0\Sigma^K_{\gamma\Bar{\gamma}}(\vect{k};\omega)G^A_{\gamma\Bar{\gamma}}(\vect{k};\omega)_0 \\
    &=\frac{-i4\pi}{(\omega-\omega_\gamma(\vect{k}))^2+\varepsilon^2}
    \sum_{\eta=\pm 1}\big\{K_{1\gamma}\coth\left[\tilde{\beta} E_\gamma(\vect{k}+ \eta\vect{q})/2\right]\delta(\omega+ \eta\tilde{\omega}_n(\vect{q}) -\omega_\gamma(\vect{k}+\eta\vect{q})) \\
    &+K_{2\gamma}\coth\left[\tilde{\beta} E_\gamma(-\vect{k}-\eta\vect{q})/2\right]\delta(\omega+\eta\tilde{\omega}_n(\vect{q})+\omega_{\underline{\gamma}}(-\vect{k}-\eta\vect{q}))\big\}.
\end{split}
\end{equation}
Here, $\omega_{\underline{\alpha}}=\omega_\beta, \omega_{\underline{\beta}}=\omega_\alpha$ and
\begin{equation}
\begin{split}
K_{1\alpha}& =\frac{\abs{A_n(\vect{q})}^2}{8\hbar^2} | \Omega_{\bar{\alpha}\alpha}(n,\eta\vect{q},\vect{k}) |^2  \text{, \;\;} K_{2\alpha} =\frac{\abs{A_n(\vect{q})}^2}{8\hbar^2} | \Omega_{\alpha\beta}(n,\eta\vect{q},\vect{k}) |^2 \\
K_{1\beta}& =\frac{\abs{A_n(\vect{q})}^2}{8\hbar^2} | \Omega_{\bar{\beta}\beta}(n, -\eta\vect{q},-\vect{k} ) |^2  \text{, \;\;} K_{2\beta} =\frac{\abs{A_n (\vect{q})}^2}{8\hbar^2} | \Omega_{\bar{\alpha}\bar{\beta}}(n,-\eta\vect{q},-\vect{k}) |^2.
\end{split}
\end{equation}
Performing the frequency integral in Eq.~(11) of the main text yields the distribution functions in Eq.~(13).
To regularize the contributions from resonant processes, we phenomenologically introduce in Eq.~\eqref{eq:KeldyshComponent} a finite damping rate $\varepsilon=\alpha_G \omega_\gamma (0)$ parametrized by the Gilbert damping $\alpha_G$.

%%%%%%
\section{Material parameters and linewidth sensitivity}
Our calculations are based on a minimal model designed to capture the essential symmetry-governed features of altermagnets rather than to provide a first-principles parametrization of a specific material. Below, we discuss the physical basis for the material parameters used in the calculations and the sensitivity of the results to the magnon linewidth.

The values $J_1=25.7$~meV and $a=4$~\AA\ set the characteristic energy and length scales of the model. These values are comparable to those reported for $\alpha$-MnTe, where the dominant exchange interaction is approximately $19$--$23$~meV and the in-plane lattice constant is $4.15$~\AA~\cite{Hermann_2019}. The temperature $T=298$~K follows from the dimensionless choice $\beta J_1=1$. The ratios $\bar{J}_2/J_1$, $\Delta J_2/J_1$, and $K_z/J_1$ were chosen such that the model supports a stable collinear $d$-wave altermagnetic state with a clearly resolved magnon spin splitting.

The nearest-neighbor exchange-striction coefficient, $\alpha_1=-53$~meV/\AA, was chosen to be representative of experimentally and theoretically established magnon--phonon coupling strengths. Its magnitude is comparable to the exchange-striction strength of approximately $65$~meV/\AA\ extracted from the measured magnon--phonon spectrum of the triangular-lattice antiferromagnet LiCrO$_2$, and to $\alpha_1\approx-53$~meV/\AA\ obtained from first-principles calculations for NiO~\cite{Gopal_2017}. Although LiCrO$_2$ and NiO have lattice structures different from the Lieb lattice considered here, these values provide a representative scale for bond-length-induced exchange modulation. We further set $\bar{\alpha}_2/\alpha_1=\bar{J}_2/J_1$ and $\Delta\alpha_2/\alpha_1=\Delta J_2/J_1$, such that the relative strengths of the exchange-striction channels follow those of the corresponding equilibrium exchange interactions.

The coherent phonon amplitude corresponds to an atomic displacement of approximately $4$~pm, consistent with the few-picometer displacements reported for coherently driven optical phonons in solids. Representative examples include an estimated displacement of approximately $5$~pm in bilayer graphene~\cite{Gierz_prl2015} and an initial displacement of $2.7$~pm for the $A_{1g}$ mode in bismuth~\cite{Johnson_prb2013}. The phonon frequencies $\omega_0^{(2)}$ were selected to access the distinct single- and two-magnon resonance regimes identified in Eq.~(13) of the main text, rather than to reproduce the phonon spectrum of a particular material.

Gilbert damping was introduced phenomenologically to account for the finite magnon lifetime and enters through the linewidths of the magnon propagators via the damping rate $\varepsilon=\alpha_G\omega_\gamma(0)$. Its primary effect is therefore to modify the width and amplitude of the resonant response. To assess the sensitivity of the spin current to the linewidth, we have repeated the calculations for $\alpha_G=0.05$ and $\alpha_G=0.20$, keeping all other parameters fixed, to bracket the value $\alpha_G=0.10$ used in the main text. This was done for all three phonon spectra in Fig.~3 of the main text, and the results are shown in Fig. S2.
 
In all cases, increasing the linewidth reduces the magnitude of the spin current. However, the spin current retains its $d$-wave symmetry with respect to the phonon momentum, and the low-frequency and high-frequency channels retain opposite signs along the altermagnetic directions. In the intermediate regime, where the current reverses as a function of $\abs{\vect{q}}$ at a fixed propagation direction, the reversal persists and the switching boundary in $\vect{q}$ space is located at the same position for all values of $\alpha_G$. This is expected, since the switching is governed by the resonance conditions, which are set by the magnon and phonon dispersions, whereas the linewidth primarily controls the strength of the resonances. Changing the constant $\alpha_G$ therefore leads only to a quantitative broadening and rescaling of the spin current, while the central results shown in Fig. 3 of the main text remain unaffected.

\begin{figure}[ht] 
\centering 
\includegraphics[width=1\linewidth]{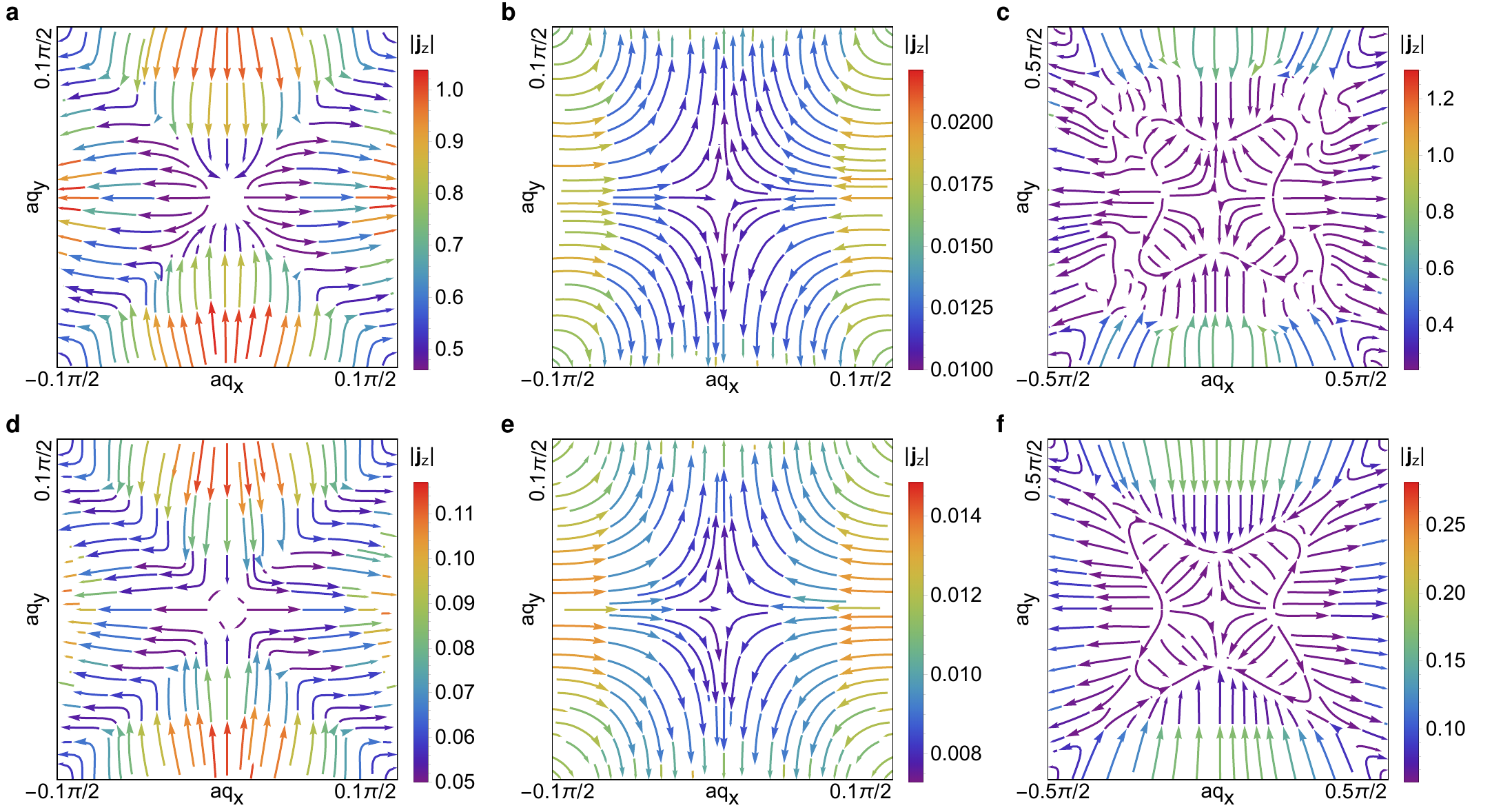}  
\caption{ Dimensionless spin current $\vect{j}^z(\vect{q})=(4\pi^2 a/J_1)\vect{J}^z(\vect{q})$ as a function of the phonon momentum $\vect{q}$ for three phonon spectra $\tilde{\omega}_2(\vect{q})=\omega_0^{(2)}\sqrt{1-a^2(q_x^2+q_y^2)/8}$ and two values of Gilbert damping: (\textbf{a})--(\textbf{c}) $\alpha_G = 0.05$ and (\textbf{d})--(\textbf{f}) $\alpha_G= 0.20$. The phonon frequency is $\omega_0^{(2)}=0.1\omega_0$ in (\textbf{a}),(\textbf{d}), $2.0\omega_0$ in (\textbf{b}),(\textbf{e}), and $1.3\omega_0$ in (\textbf{c}),(\textbf{f}). All other parameters are as in Fig.~3 of the main text.}
\label{Sfig:2}
\end{figure}

\end{widetext}

%%% References %%%%%%%%%%%%%%%%%%%%%%%%%%%%%%%%%%%%%%%%%%%%%%%%%%%%%%%%%%%% 


\begin{thebibliography}{99}
\bibitem{Comment1}In the absence of SOC, the phonon system is independent of the spin configuration, causing the time reversal operator $\mathcal{T}$ in the magnetic plane point group to be irrelevant. Consequently, in Eq.~\eqref{eq:SymRels},  $\boldsymbol{S} \in \{E,C_{2z}, \sigma_x, \sigma_y, C_{4z}^+, C_{4z}^-, \sigma_{xy}, \sigma_{x\Bar{y}} \}$.
\bibitem{comment4}For simplicity, we assume radially isotropic exchange parameters, which provide a good microscopic approximation for direct-exchange magnets with quenched orbital degrees of freedom.
We further neglect SOC-mediated magnetoelastic interactions.
\bibitem{comment3}Note that the phonon angular momentum $\vect{J} \propto \vect{q} \times \hat{\boldsymbol{z}}$ does not enter $H_I$ explicitly.  The coupling is governed by the projections  $\hat{\vect{n}}_{ij}\!\cdot\!\Delta\vect{u}_{ij}$, 
which are insensitive to circular atomic motion in planes containing $\hat{\boldsymbol{z}}$ because $\hat{\vect{n}}_{ij}\!\cdot\!\hat{\boldsymbol{z}} = 0$. 
Consequently, rotational components of the lattice displacements do not couple directly to the magnon sector.
Phonon modes with a pronounced longitudinal bond-stretching character give rise to the strongest magnon--phonon coupling.
\end{thebibliography}
\end{document}